\documentclass[conference,a4paper]{IEEEtran}
\IEEEoverridecommandlockouts
\usepackage[utf8]{inputenc} 
\usepackage[T1]{fontenc}
\usepackage{url}              
\usepackage{cite}             

\usepackage[cmex10]{amsmath}  
\usepackage{mleftright}       
\mleftright                   

\usepackage{graphicx}         
\usepackage{booktabs}         

\usepackage{amsmath}
\usepackage{amsthm}
\usepackage{cases}
\usepackage{amsfonts}
\usepackage{amssymb}
\usepackage{enumerate}
\usepackage{float}
\usepackage{caption}
\usepackage{verbatim}
\usepackage{dblfloatfix}
\usepackage[cmintegrals]{newtxmath}
\usepackage[ruled,linesnumbered]{algorithm2e}
\usepackage[numbers,sort&compress]{natbib}

\newtheorem{theorem}{Theorem}
\newtheorem{lemma}{Lemma}
\newtheorem{definition}{Definition}

\newtheorem{proposition}{Proposition}
\newtheorem{example}{Example}
\begin{document}

\title{On Strong Secrecy for Multiple Access Channel with States and causal CSI} 


%
%

%
 \author{%
      \IEEEauthorblockN{Yiqi Chen\IEEEauthorrefmark{1}\IEEEauthorrefmark{2},
                      Tobias Oechtering\IEEEauthorrefmark{2},
                      Mikael Skoglund\IEEEauthorrefmark{2},
                      and Yuan Luo\IEEEauthorrefmark{1}}
    \IEEEauthorblockA{\IEEEauthorrefmark{1}%
                    Shanghai Jiao Tong University,  
                    Shanghai,
                    China, 
                    \{chenyiqi,yuanluo\}@sjtu.edu.cn}
   \IEEEauthorblockA{\IEEEauthorrefmark{2}%
                    KTH Royal Institute of Technology,  
                    100 44 Stockholm,
                    Sweden, 
                    \{oech,skoglund\}@kth.se}
    \thanks{This work was supported in part by National Key R\&D Program of China under Grant 2022YFA1005000, and National Natural Science Foundation of China under Grant 61871264 and China Scholarship Council (CSC).} 
}

\maketitle

\begin{abstract}
  Strong secrecy communication over a discrete memoryless state-dependent multiple access channel (SD-MAC) with an external eavesdropper is investigated. The channel is governed by discrete memoryless and i.i.d. channel states and the channel state information (CSI) is revealed to the encoders in a causal manner. An inner bound of the capacity is provided. To establish the inner bound, we investigate coding schemes incorporating wiretap coding and secret key agreement between the sender and the legitimate receiver. Two kinds of block Markov coding schemes are studied. The first one uses backward decoding and Wyner-Ziv coding and the secret key is constructed from a lossy reproduction of the CSI. The other one is an extended version of the existing coding scheme for point-to-point wiretap channels with causal CSI. We further investigate some capacity-achieving cases for state-dependent multiple access wiretap channels (SD-MAWCs) with degraded message sets. It turns out that the two coding schemes are both optimal in these cases.
\end{abstract}

\section{Introduction}
\label{sec:Introduction}

Secure communication over a discrete memoryless channel (DMC) was first studied in \cite{wyner1975wire} where the sender communicates to the legitimate receiver over the main channel in the presence of an external eavesdropper through a degraded version of the main channel. The model was further extended to a more general case called broadcast channels with confidential messages in \cite{csiszar1978broadcast}. Following these landmark papers, secrecy capacity results of different generalized MAC models have been reported in recent years \cite{liang2008multiple,yassaee2010multiple,wiese2013strong}.

The point-to-point channel with causal channel state information (CSI) at the encoder was studied in \cite{shannon1958channels}. The optimal coding scheme in \cite{shannon1958channels}, called Shannon strategy, was proved to be suboptimal for state-dependent multiple access channels (SD-MACs) in \cite{lapidoth2012multiple}, where block Markov coding with Wyner-Ziv coding and backward decoding achieved a strictly larger achievable region in some cases. SD-MAC with independent states at each sender was investigated in \cite{lapidoth2012multiple2}. 

Wiretap channels with noncausal CSI at the encoder was studied in \cite{chen2008wiretap} and lower and upper bounds of the secrecy capacity were presented by combining Gel'fand-Pinsker (GP) coding \cite{gel1980coding} and wiretap coding. The model was further studied in \cite{goldfeld2019wiretap} with a more stringent secrecy constraint. 
For a state-dependent channel, knowing CSI can help the secure transmission over the channel, since the system participants have some additional resources to construct a secret key and encrypt part of messages by using Shannon's one-time pad cypher \cite{shannon1949communication}. This combination of wiretap codes and secret key agreement has been used in recent works about state-dependent wiretap channels (SD-WTCs). In \cite{chia2012wiretap}, Chia and El Gamal addressed the secure communication over wiretap channels with causal CSI at both the encoder and the decoder using block Markov coding.  Wiretap channels with causal CSI at encoder was studied in  \cite{fujita2016secrecy} and \cite{sasaki2019wiretap}. The results were further strengthened in \cite{sasaki2021wiretap} to semantic secrecy constraint by using the soft-covering method \cite{goldfeld2019wiretap}.

In this paper, we consider SD-MACs with causal CSI at encoders with an external eavesdropper and a strong secrecy constraint. 
We design a new coding scheme which incorporates block Markov coding with Wyner-Ziv coding, backward decoding, and secret key agreement \cite{csiszar2000common}. Our inner bound includes previous results and recovers existing works for other models as special cases.

The rest of the paper is organized as follows: In Section II, we give the needed notations and definitions of the channel model considered in this paper. In Section III, we give the inner bound result, which consists of three different regions, each corresponding to a different coding scheme. The first two coding schemes (we refer them as Coding scheme 1 and Coding scheme 2, respectively) combine wiretap coding and secret key agreement while the third coding scheme only uses wiretap coding. The first coding scheme is presented in Section IV. Section V provides some numerical examples and a capacity achieved case that both Coding schemes 1 and 2 are optimal. Section VI concludes this paper.

\begin{figure}[t]
  \includegraphics[scale=0.5]{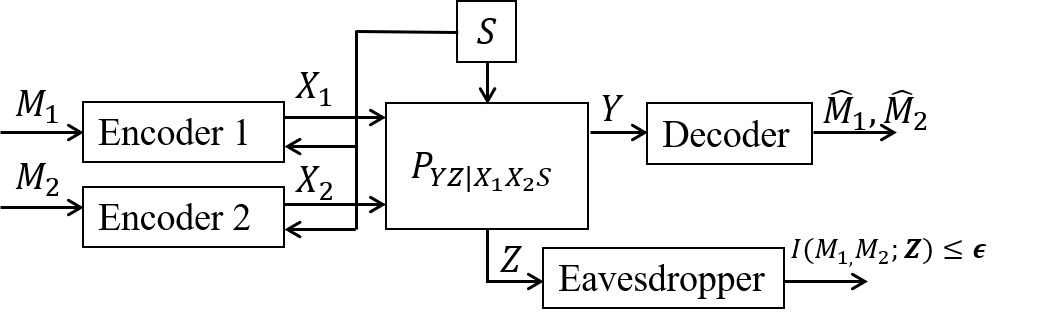}
  \caption{SD-MAWC with Causal CSI at Encoders.}
\end{figure}

\section{notations and definitions}
\label{sec:definitions}

Throughout this paper, random variables and sample values are denoted by capital letters and lowercase letters, e.g. $X$ and $x$. Sets are denoted by calligraphic letters. Capital and lowercase letters in boldface represents n-length random and sample sequences, respectively, e.g. $\boldsymbol{X}=(X_1,X_2,\dots,X_n)$ and $\boldsymbol{x}=(x_1,x_2,\dots,x_n)$. Let $\mathcal{X}^n$  be the n-fold Cartesian product of $\mathcal{X}$, which is the set of all possible $\boldsymbol{x}$. To denote substrings, let $X^{b}=(X_1,X_2,\dots,X_b)$ and $X^{[b+1]}=(X_{b+1},X_{b+2},\dots,X_n)$. We use $P_X$ to denote the probability mass functions (PMFs) of random variable $X$ and $P_{XY},P_{X|Y}$ to denote the joint PMFs and conditional PMFs, respectively. The set of all PMFs over a given set $\mathcal{X}$ is written by $\mathcal{P}(\mathcal{X})$. For a sequence $\boldsymbol{x}$ generated i.i.d. according to some distribution $P_X$, we denote $P_X^n(\boldsymbol{x})=\prod_{i=1}^n P_X(x)$.

\begin{definition}\label{def: channel model}
  A discrete memoryless multiple access wiretap channel with state is defined by a stochastic transition matrix $\mathcal{W}:\mathcal{X}_1\times\mathcal{X}_2\times\mathcal{S}\hspace{-0.05in}\to\hspace{-0.05in}\mathcal{Y}\times\mathcal{Z}$, where $\mathcal{X}_1,\mathcal{X}_2$ are finite input alphabets, $\mathcal{Y}, \mathcal{Z}$ are finite output alphabets, $\mathcal{S}$ is a finite state alphabet. The transition probability from input sequences $(\boldsymbol{x}_1,\boldsymbol{x}_2)$ to the output sequences $(\boldsymbol{y},\boldsymbol{z})$ given state sequence $\boldsymbol{s}$ is 
  \begin{align*}
    P_{YZ|X_1X_2S}(\boldsymbol{y},\boldsymbol{z}|\boldsymbol{x}_1,\boldsymbol{x}_2,\boldsymbol{s}) = \prod_{i=1}^n P_{YZ|X_1X_2S}(y_i,z_i|x_{1i},x_{2i},s_i),
  \end{align*}
  where $\boldsymbol{y}$ is the output of the main channel and $\boldsymbol{z}$ is the output of the wiretap channel. The channel state sequences are generated by a discrete memoryless source such that $P_S^n(\boldsymbol{s})=\prod_{i=1}^n P_S(s_i)$.
\end{definition}

\begin{definition}
  A  code $(2^{nR_1},2^{nR_2},f^n_1,f^n_2,\phi)$ consists of message sets $\mathcal{M}_1=[1:2^{nR_1}],\mathcal{M}_2=[1:2^{nR_2}]$, sets of encoders $f_{i,j}:\mathcal{M}_i \times \mathcal{S}^j \to \mathcal{X}_i, i=1,2,j=1,\dots,n$ and a decoder $\phi:\mathcal{Y}^n\to\mathcal{M}_1\times\mathcal{M}_2$. To transmit message $M_i$, the sender sends codeword $\boldsymbol{X}_i$ with each component $X_{ij}$ generated according to $f_{i,j}(M_i,\boldsymbol{S}^j)$. The receiver receives $\boldsymbol{Y}$ and estimates the message by the decoder $(\hat{M}_1,\hat{M}_2)=\phi(\boldsymbol{Y})$. The average decoding error of a code is 
  \begin{align*}
    &P_e = \frac{1}{|\mathcal{M}_1||\mathcal{M}_2|}\sum_{m_1\in\mathcal{M}_1}\sum_{m_2\in\mathcal{M}_2}\\
    &\quad\quad\quad\quad\quad Pr\{(M_1,M_2)\neq(\hat{M}_1,\hat{M}_2)|M_1=m_1,M_2=m_2\}.
  \end{align*}
The information leakage of the code is defined as 
\begin{align*}
  I(M_1,M_2;\boldsymbol{Z}),
\end{align*}
where $\boldsymbol{Z}$ is the output of the wiretap channel.
\end{definition}
\begin{definition}
  A rate pair $(R_1,R_2)$ is said to be achievable for the multiple access wiretap channel with causal channel state information at encoders if for any $\epsilon>0$, there exists an $n_0$ such that for all $n>n_0$ there exists a code $(2^{nR_1},2^{nR_2},f^n_1,f^n_2,\phi)$ with
  \begin{align*}
    P_e \leq \epsilon,I(M_1,M_2;\boldsymbol{Z}) \leq \epsilon.
  \end{align*}
 The secrecy capacity of the multiple access wiretap channel is the closure of all sets of achievable rate pairs.
\end{definition}
%
\section{Main RESULTS}\label{sec: main results}
Given $P_S$ and $P_{YZ|X_1X_2S}$, define region $\mathcal{R}_{1i}(P_{VUU_1U_2X_1X_2|S}),i=1,2,3$ as sets of real number pairs $(R_1,R_2)$ such that

\begin{align*}
  &\mathcal{R}_{11}(P_{VUU_1U_2X_1X_2|S})=\\
  &\left\{
    \begin{aligned}
      &R_1 \leq \min\{I(U_1;Y|V,U,U_2)-I(U_1;Z|S,U)+R_{11},\\
      &\quad\quad\quad\quad\quad\quad\quad\quad\quad\quad\quad\quad I(U_1;Y|V,U,U_2)\};\\
      &R_2 \leq \min\{I(U_2;Y|V,U,U_1)-I(U_2;Z|S,U)+R_{21},\\
      &\quad\quad\quad\quad\quad\quad\quad\quad\quad\quad\quad\quad I(U_2;Y|V,U,U_1)\};\\  
      &R_1 + R_2 \leq \min\{R_{SUM},R_{SUM}-I(U_1,U_2;Z|S,U)\\
      &\quad\quad\quad\quad\quad\quad\quad\quad\quad\quad\quad\quad\quad +R_{11}+R_{21}\};\\
      &R_{11}+R_{21} \leq I(V;Y)-I(V;U,Z).
    \end{aligned}
  \right.
\end{align*}
\begin{align*}
  \begin{split}
    &\mathcal{R}_{12}(P_{VUU_1U_2X_1X_2|S})=\\
    &\left\{
    \begin{aligned}
      &R_1 \leq \min\{I(U_1;Y|V,U,U_2),R_{11}\};\\
      &R_2 \leq \min\{I(U_2;Y|V,U,U_1)-I(U_2;Z|S,U,U_1)+R_{21},\\
      &\quad\quad\quad\quad\quad\quad\quad\quad\quad\quad\quad\quad I(U_2;Y|V,U,U_1)\};\\
      &R_1 + R_2 \leq \min\{R_{SUM},R_{SUM} - I(U_2;Z|S,U,U_1) + R_{21}, \\
      &\quad\quad\quad I(U_2;Y|V,U,U_1)-I(U_2;Z|S,U,U_1)+R_{11}+R_{21}\};\\
      &R_{11}+R_{21}\leq I(V;Y)-I(V;Z,U,U_1),
    \end{aligned}
  \right. \\
\end{split}
\end{align*}
\begin{align*}
  \begin{split}
  &\mathcal{R}_{13}(P_{VUU_1U_2X_1X_2|S})=\\
  &\left\{
    \begin{aligned}
      &R_1 \leq \min\{I(U_1;Y|V,U,U_2)-I(U_1;Z|S,U,U_2)+R_{11},\\
      &\quad\quad\quad\quad\quad\quad\quad\quad\quad\quad\quad\quad I(U_1;Y|V,U,U_2)\};\\
      &R_2 \leq \min\{I(U_2;Y|V,U,U_1),R_{21}\};\\
      &R_1 + R_2 \leq \min\{R_{SUM},R_{SUM} - I(U_1;Z|S,U,U_2) + R_{11}, \\
      &\quad\quad\quad I(U_1;Y|V,U,U_2)-I(U_1;Z|S,U,U_2)+R_{11}+R_{21}\};\\
      &R_{11}+R_{21}\leq I(V;Y)-I(V;Z,U,U_2),
    \end{aligned}
  \right.\\
  \end{split}
\end{align*}
where $R_{SUM}=\min\{I(U_1,U_2;Y|V,U),I(V,U,U_1,U_2;Y)-I(V;S)\}$ and joint distribution such that $P_SP_{VUU_1U_2X_1X_2YZ|S}=P_{S}P_{V|S}P_{U}P_{U_1|U}P_{U_2|U}$ $P_{X_1|UU_1S}P_{X_2|UU_2S}P_{YZ|X_1X_2S}$.

 Let $\mathcal{R}_1$ be the convex hull of 
\begin{align*}
  \bigcup_{\substack{P_{V|S}P_{U}P_{U_1|U}P_{U_2|U}\\P_{X_1|UU_1S}P_{X_2|UU_2S}}} \bigcup_{i=1}^3\mathcal{R}_{1i}(P_{VUU_1U_2X_1X_2|S}).
\end{align*}

Removing the secrecy constraint, region $\mathcal{R}_1$ reduces to \cite[Theorem 3]{lapidoth2012multiple}. Further define $\mathcal{R}_{2i}(P_{U_1U_2X_1X_2|S}),i=1,2,3$, as sets of real number pairs $(R_1,R_2)$ such that
\begin{align*}
  &\mathcal{R}_{21}(P_{U_1U_2X_1X_2|S})=\\
  &\left\{
    \begin{aligned}
      &R_1 \leq \min\{I(U_1;Y|U_2)-I(U_1;Z|S)+R_{11},I(U_1;Y|U_2)-R_{12}\};\\
      &R_2 \leq \min\{I(U_2;Y|U_1)-I(U_2;Z|S)+R_{21},I(U_2;Y|U_1)-R_{22}\};\\ 
      &R_1+R_2 \leq \min\{I(U_1,U_2;Y)-I(U_1,U_2;Z|S)+H(S|Z)\\
      &\quad\quad\quad\quad -H(S|U_1U_2Y),I(U_1,U_2;Y)-H(S|U_1U_2Y)\};\\
      &R_{11}+R_{12}+R_{21}+R_{22} < H(S|Z);\\
      &R_{12}+R_{22} > H(S|Y,U_1,U_2).
    \end{aligned}
  \right.
\end{align*}
\begin{align*}
  \begin{split}
    &\mathcal{R}_{22}(P_{U_1U_2X_1X_2|S})=\\
    &\left\{
    \begin{aligned}
      &R_1 \leq \min\{R_{11}, I(U_1;Y|U_2)-R_{12}\};\\
      &R_2 \leq \min\{I(U_2;Y|U_1)-I(U_2;Z|S,U_1)+R_{21},\\
      &\quad\quad\quad\quad\quad I(U_2;Y|U_1)-R_{22}\};\\  
      &R_1 + R_2 \leq \min\{I(U_2;Y|U_1)-I(U_2;Z|S,U_1)+H(S|Z,U_1)\\
      &\quad\quad\quad\quad\quad -H(S|U_1U_2Y),I(U_1,U_2;Y)-H(S|U_1U_2Y)\};\\
      &R_{11}+R_{12}+R_{21}+R_{22} < H(S|Z);\\
      &R_{12}+R_{22} > H(S|Y,U_1,U_2).
    \end{aligned}
  \right.\\
\end{split}
\end{align*}
\begin{align*}
  \begin{split}
  &\mathcal{R}_{23}(P_{U_1U_2X_1X_2|S})=\\
  &\left\{
    \begin{aligned}
      &R_1 \leq \min\{I(U_1;Y|U_2)-I(U_1;Z|S,U_2)+R_{11},\\
      &\quad\quad\quad\quad\quad I(U_1;Y|U_2)-R_{12}\};\\
      &R_2 \leq \min\{R_{21},I(U_2;Y|U_1)-R_{22}\};\\  
      &R_1 + R_2 \leq \min\{I(U_1;Y|U_2)-I(U_1;Z|S,U_2)+H(S|Z,U_2)\\
      &\quad\quad\quad\quad\quad -H(S|U_1U_2Y),I(U_1,U_2;Y)-H(S|U_1U_2Y)\};\\
      &R_{11}+R_{12}+R_{21}+R_{22} < H(S|Z);\\
      &R_{12}+R_{22} > H(S|Y,U_1,U_2),
    \end{aligned}
  \right.
  \end{split}
\end{align*}
where the joint distribution $P_SP_{U_1U_2X_1X_2YZ|S}=P_SP_{U_1}P_{U_2}P_{X_1|U_1S}P_{X_2|U_2S}P_{YZ|X_1X_2S}$. Similarly, let $\mathcal{R}_2$ be the convex hull of
\begin{align*}
  \bigcup_{\substack{P_{U_1}P_{U_2}\\P_{X_1|U_1S}P_{X_2|U_2S}}} \bigcup_{i=1}^3\mathcal{R}_{2i}(P_{U_1U_2X_1X_2|S}).
\end{align*}
Further define region $\mathcal{R}_3$ by setting $R_{11}=R_{21}=0$ in $\mathcal{R}_1$ and removing $S$ in all the subtraction terms related to $Z$ $(I(U_1;Z|S,U), I(U_1,U_2;Z|S,U), etc.)$ from $\mathcal{R}_1$ with joint distribution $P_SP_{VUU_1U_2X_1X_2YZ|S}=P_{S}P_{V|S}P_{U}P_{U_1|U}P_{U_2|U}$ $P_{X_1|UU_1S}P_{X_2|UU_2S}P_{YZ|X_1X_2S}$. 

Throughout the paper, we refer coding scheme of region $\mathcal{R}_1$ as Coding scheme 1 and coding scheme of region $\mathcal{R}_2$ as Coding scheme 2.
\begin{theorem}\label{the: mawc with casual csi}
  Let $\mathcal{R}^{CSI}$ be the convex hull of $\mathcal{R}_1 \cup \mathcal{R}_2 \cup \mathcal{R}_3$.
  The secrecy capacity of multiple access wiretap channels with causal channel state information $C^{CSI}$ satisfies
  \begin{align*}
    \mathcal{R}^{CSI} \subseteq C^{CSI}.
  \end{align*}
\end{theorem}
In Region $\mathcal{R}_1$, auxiliary random variable $U$ is the common message transmitted by the senders to help the decoder find the lossy description $V$ of the state from last block since both of them observe the state sequence. Auxiliary random variable $V$ also contains information about the secret key used in the coding scheme. The coding scheme for region $\mathcal{R}_1$ is given in Section \ref{sec: coding scheme for region r1}. We omit coding schemes for regions $\mathcal{R}_2$ and $\mathcal{R}_3$ since the coding scheme for region $\mathcal{R}_2$ is an extension of that in \cite[Section III]{sasaki2019wiretap} by splitting both the Slepian-Wolf coding index and secret key into two independent parts and coding scheme for $\mathcal{R}_3$ follows by skipping the \emph{Key Message Codebook Generation} and setting $R_{11}=R_{21}=0$ in \emph{Message Codebook Generation} in Coding scheme 1 provided in Section \ref{sec: coding scheme for region r1}.

\section{Coding scheme for region $\mathcal{R}_1$}\label{sec: coding scheme for region r1}
In this section we give the coding scheme for region $\mathcal{R}_{1}$.  For the secrecy part, the main difference between the coding scheme below and that in \cite{sasaki2019wiretap} is that instead of constructing the secret key from state sequences directly, we construct the secret key from a lossy description of the state sequence. 

The coding scheme with backward decoding is divided into $B+1$ blocks. In block $2\leq b \leq B+1$, upon observing the channel state sequence $\boldsymbol{s}_{b-1}$ in the last block $b-1$, the encoder finds a lossy description of the state sequence $\boldsymbol{v}_{b-1}$ and constructs a secret key from $\boldsymbol{v}_{b-1}$. In Blocks 1 and $B+1$, no meaningful message is transmitted. We further assume that $I(U_1,U_2;Y)>\mu>0$ for some positive number $\mu$, otherwise, as shown by \cite[(73b)]{lapidoth2012multiple}, the rates of both senders are zero. The codeword length of blocks 1 to $B$ is $n$ and the length of the last block is 
\begin{align}
  \label{eq: codeword length of last block}\widetilde{n}= n\frac{R_0}{\mu},
\end{align}
where $R_0>0$ is a positive real number such that $2^{nR_0}$ is the size of the auxiliary message codebook defined in following paragraphs. The last block is only used to transmit the information of state sequence $\boldsymbol{s}_{B}$ and the setting of $\widetilde{n}$ enables the decoder to decode the information correctly. Assume $I(V;Y)-I(V;Z)>\tau_1> 0$ with Markov chain relation $V-S-(Y,Z)$. Otherwise, the secret key rate is 0 and the resulting achievable region is $\mathcal{R}_{3}$. The senders cooperate with each other since both of them observe the same channel states, and send a common message to convey the lossy description of the state sequence they observe. Based on this lossy description of the channel state, the decoder is able to decode more messages and this leads to a larger achievable rate region.

The backward decoding follows the principles in \cite[Appendix G]{lapidoth2012multiple} with an additional secret key decryption due to the secret key agreement. The decoder first decodes Block $B+1$ to get the index of the state lossy description of Block $B$. In the subsequent decoding blocks, the decoder first finds the lossy description of the current state sequence and then performs joint typical decoding to decode the common and private messages with the help of this lossy description.

Here, we present an intuitive explanation of the secret key agreement in the coding scheme. For simplicity, we only consider the encryption and decryption operations of Sender 1. In each block $b, b\in[2:B]$, upon observing the channel state sequence $\boldsymbol{s}_{b-1}$ and its lossy description $\boldsymbol{v}_{b-1}$, the sender generates a secret key $k_{1,b}$ by a mapping $\kappa: \mathcal{T} \to \mathcal{K}_1$, where $\mathcal{T}$ is the index set of $\boldsymbol{V}$ in Wyner-Ziv subcodebooks. This key is used to encrypt the message $m_{1,b}$ by computing $c_{1,b}=m_{1,b} \oplus k_{1,b}$. In the backward decoding step block $b, b\in[1:B]$, the decoder has the knowledge about the Wyner-Ziv index $k_{0,b+1}$ from the last block $b+1$. Now it is possible to find $\boldsymbol{v}_{b}$, which is the lossy description of the state sequence $\boldsymbol{s}_b$ in current block. The decoder then reproduces the secret key $k_{1,b+1}$ used for block $b+1$ and decrypts the message $m_{1,b+1}$ by computing $m_{1,b+1}=c_{1,b+1} \ominus k_{1,b+1}$.


Given $P_S$ and channel $P_{YZ|X_1X_2S}$, let $R_0, \widetilde{R}_1,\widetilde{R}_2,R_{10},$ $R_{11},R_{20},R_{21}, R_{K_1}$ be positive real numbers with constraints
\begin{align*}
  R_0 \geq I(V;S)-I(V;Y),&\\
  \widetilde{R}_1 \leq I(U_1;Y|V,U,U_2),\widetilde{R}_2 \leq I(U_2;Y|V,U,U_1),&\\
  \widetilde{R}_1 - R_{10} > I(U_1;Z|S,U),\widetilde{R}_2 - R_{20} > I(U_2;Z|S,U),&\\
  \widetilde{R}_1 + \widetilde{R}_2 - R_{10} - R_{20} > I(U_1,U_2;Z|S,U),&\\
  R_{K_1}=R_{11} + R_{21} \leq I(V;Y) - I(V;Z),&\\
  R_1 = R_{10}+R_{11}, R_{2}=R_{20}+R_{21}&
\end{align*}
under fixed joint distribution $$P_{S}P_{V|S}P_UP_{U_1|U}P_{U_2|U}P_{X_1|UU_1S}P_{X_2|UU_2S}P_{YZ|X_1X_2S}.$$

\emph{Key Message Codebook Generation: } Given $\tau>0$ and $\delta>0$, let $R_K=I(V;S)+\tau$. In each block $2\leq b \leq B+1$, the sender generates a codebook $\mathcal{C}_{K_b}=\{\boldsymbol{v}(l)\}_{l=1}^{2^{nR_{K}}}$ consists of $2^{nR_K}$ codewords, each i.i.d. generated according to distribution $P_V$ such that $P_V(v)=\sum_{s\in\mathcal{S}}P_S(s)P_{V|S}(v|s)$ for any $v\in\mathcal{V}$. Partition the codebook $\mathcal{C}_{K_b}$ into $2^{nR_{K_0}}$ sub-codebooks $\mathcal{C}_{K_b}(k_{0,b})$, where $k_{0,b}\in[1:2^{nR_{K_0}}]$ and $R_{K_0}=I(V;S)-I(V;Y)+2\tau$. Let $\mathcal{T}$ be the index set of codewords in each subcodebook $\mathcal{C}_{K_b}(k_{0,b})$ such that $|\mathcal{T}|=|\mathcal{C}_{K_b}(k_{0,b})|$ for any $k_{0,b}\in[1:2^{nR_{K_0}}]$. For each codebook $\mathcal{C}_{K_b},$ construct a secret key mapping $\kappa:\mathcal{T}\to [1:2^{nR_{K_1}}].$ Denote the resulted secret key by $K_{1,b}$.

\emph{Auxiliary Message Codebook Generation: } For each block $b$, generate auxiliary message codebook $\mathcal{C}_b=\{\boldsymbol{u}(m_0)\}_{m_0=1}^{2^{nR_0}}$ i.i.d. according to distribution $P_{U}$, where $R_0=I(V;S)-I(V;Y)+3\tau$.

\emph{Message Codebook Generation:}  
Block $b, b\in[1:B]$: For each $m_0$, generate codebook $\mathcal{C}_{1,b}(m_0)=\{\boldsymbol{u}_1(m_0,l)\}_{l=1}^{2^{n\widetilde{R}_1}}$ containing $2^{n\widetilde{R}_1}$ codewords, each i.i.d. generated according to distribution $P_{U_1|U}$. Partition each $\mathcal{C}_{1,b}(m_0)$ into $2^{nR_{10}}$ subcodebooks $\mathcal{C}_{1,b}(m_0,m_{10})$, where $m_{10}\in[1:2^{nR_{10}}].$ For each subcodebook $\mathcal{C}_{1,b}(m_0,m_{10})$, partition it into two-layer subcodebooks $\mathcal{C}_{1,b}(m_0,m_{10},m_{11})=\{\boldsymbol{u}_1(m_0,m_{10},m_{11},l_1)\}_{l_1=1}^{2^{nR_1'}},$ where $m_{11}\in[1:2^{nR_{11}}],R_1' := \widetilde{R}_1-R_{10}-R_{11}$. Likewise, generate codebook $\mathcal{C}_{2,b}=\{\boldsymbol{u}_2(m_0,l)\}_{l=1}^{2^{n\widetilde{R}_2}}$ with codewords i.i.d. generated according to $P_{U_2|U}$, and then partition it into two-layer sub-codebooks $\mathcal{C}_{2,b}(m_0,m_{20},m_{21})=\{\boldsymbol{u}_2(m_0,m_{20},m_{21},l_2)\}_{l_2=1}^{2^{nR_2'}}$, where $m_{20}\in[1:2^{nR_{20}}],m_{21}\in[1:2^{nR_{21}}],R_2':=\widetilde{R}_2-R_{20}-R_{21}$.

  Block $B+1$: For $k=1,2$, generate codebooks $\mathcal{C}_{k,B+1}$ as above with codeword length $\widetilde{n}$ defined as in \eqref{eq: codeword length of last block}.


The above codebooks are all generated randomly and independently. Denote the set of random codebooks in each block $b$ by $\bar{\textbf{C}}_b$.

\emph{Encoding: } 
 Block 1: Setting $m_{0,1}=m_{10,1}=m_{20,1}=m_{11,1}=m_{21,1}=1$, the encoder $j$ picks an index $l_j\in[1:2^{nR_j'}]$ uniformly at random, $j=1,2$. The codeword $\boldsymbol{x}_j$ is generated by $(\boldsymbol{u}(1),\boldsymbol{u}_j(1,1,1,l_j),\boldsymbol{s})$ according to $P^n_{X_j|UU_jS}(\boldsymbol{x}_j|\boldsymbol{u},\boldsymbol{u}_j,\boldsymbol{s})=\prod_{i=1}^n P_{X_j|UU_jS}(x_{ji}|u_i,u_{ji},s_i),j=1,2$. Here we omit the indices of the codewords. 

 Block $b,b\in[2:B]$: Upon observing the state sequence $\boldsymbol{s}_{b-1}$ in last block, the encoders find a sequence $\boldsymbol{v}_{b-1}$ such that $(\boldsymbol{s}_{b-1},\boldsymbol{v}_{b-1})\in T^n_{P_{SV},\delta}$ and set $m_{0,b}=k_{0,b}$, where $k_{0,b}$ is the index of subcodebook $\mathcal{C}_{k_{b-1}}(k_{0,b})$ containing $\boldsymbol{v}_{b-1}$. We also write sequence $\boldsymbol{v}_{b-1}$ as $\boldsymbol{v}(k_{0,b},t_b)$ if $\boldsymbol{v}_{b-1}$ is the $t_b$-th sequence in sub-codebook $\mathcal{C}_{K_{b-1}}(k_{0,b})$. Generate the secret key $k_{1,b}=\kappa(t_b)$ and then split it into two independent parts $(k_{11,b},k_{21,b})\in[1:2^{nR_{11}}]\times[1:2^{nR_{21}}]$. To transmit message $m_{1,b}$, Encoder $1$ splits it into two independent parts $(m_{10,b},m_{11,b})$ and computes $c_{11,b} = m_{11,b} \oplus k_{11,b} \pmod{2^{nR_{11}}}$. The encoder  selects an index $l_{1,b}\in[1:2^{nR_1'}]$ uniformly at random and generates the codeword $\boldsymbol{x}_1$ by $P^n_{X_1|UU_1S}(\boldsymbol{x}_1|\boldsymbol{u}(k_{0,b}),\boldsymbol{u}_1(k_{0,b},m_{10,b},c_{11,b},l_{1,b}),\boldsymbol{s})$.
  Likewise, the codeword $\boldsymbol{x}_2$ for Sender 2 is generated by $P^n_{X_2|UU_2S}(\boldsymbol{x}_2|\boldsymbol{u}(k_{0,b}),\boldsymbol{u}_2(k_{0,b},m_{20,b},c_{21,b},l_{2,b}),\boldsymbol{s}).$
 
  Block $B+1$: Upon observing the state sequence $\boldsymbol{s}_{B}$ in the last block, the encoder finds a sequence $\boldsymbol{v}(k_{0,B+1},t_{B+1})$ such that $(\boldsymbol{s}_{B},\boldsymbol{v}(k_{0,B+1},t_{B+1}))\in T^n_{P_{SV},\delta}$. The encoders then set $m_{0,B+1}=m_{10,B+1}=m_{11,B+1}=m_{20,B+1}=m_{21,B+1}=1$ and generate codewords $\boldsymbol{x}_1$ and $\boldsymbol{x}_2$ according to distributions $P^n_{X_1|UU_1S}(\boldsymbol{x}_1|\boldsymbol{u}(k_{0,B+1}),\boldsymbol{u}_1(k_{0,B+1},1,1,1),\boldsymbol{s})$ and $P^n_{X_2|UU_2S}(\boldsymbol{x}_2|\boldsymbol{u}(k_{0,B+1}),\boldsymbol{u}_2(k_{0,B+1},1,1,1),\boldsymbol{s})$.

\emph{Backward Decoding: } 
Block $B+1$: The decoder looks for a unique $\hat{k}_{0,B+1}$ such that $(\boldsymbol{u}_1(\hat{k}_{0,B+1},1,1,1),\boldsymbol{u}_2(\hat{k}_{0,B+1},1,1,1),\boldsymbol{y}_{B+1})\in T^n_{P_{U_1U_2Y},\delta}$.

Block $b,b\in[1:B]$: The decoder has the knowledge about $\hat{k}_{0,b+1}$ from the last block. It tries to find a unique $\boldsymbol{v}_{b}=\boldsymbol{v}(\hat{k}_{0,b+1},\hat{t}_{b+1})$ such that $(\boldsymbol{v}(\hat{k}_{0,b+1},\hat{t}_{b+1}),\boldsymbol{y}_b)\in T^n_{P_{VY},\delta}$. Using $(\hat{k}_{11,b+1},\hat{k}_{21,b+1})=\kappa(\hat{t}_{b+1})$, the decoder now computes $\hat{m}_{11,b+1}=\hat{c}_{11,b+1} \ominus \hat{k}_{11,b+1} \pmod{2^{nR_{11}}}$ and $\hat{m}_{21,b+1}=\hat{c}_{21,b+1} \ominus \hat{k}_{21,b+1} \pmod{2^{nR_{21}}}$. 
  For block $b,b\in[2:B]$, with the help of $\boldsymbol{v}(\hat{k}_{0,b+1},\hat{t}_{b+1})$, the decoder looks for a unique tuple $(\hat{k}_{0,b},\hat{m}_{10,b},\hat{m}_{20,b},\hat{c}_{11,b},\hat{c}_{21,b},\hat{l}_{1,b},\hat{l}_{2,b})$ such that 
  \begin{align*}
    &(\boldsymbol{v}_b,\boldsymbol{u}(\hat{k}_{0,b}),\boldsymbol{u}_1(\hat{k}_{0,b},\hat{m}_{10,b},\hat{c}_{11,b},\hat{l}_{1,b}),\\
    &\quad\quad\quad\quad \boldsymbol{u}_2(\hat{k}_{0,b},\hat{m}_{20,b},\hat{c}_{21,b},\hat{l}_{2,b}),\boldsymbol{y}_b)\in T^n_{P_{VUU_1U_2Y},\delta}.
  \end{align*}

  Block 1: No decoding needed since the messages transmitted in Block 1 are dummy messages. 

  The error analysis is similar to \cite[Theorem 1]{lapidoth2012multiple} and is omitted here. The security analysis is given in Appendix \ref{app: proof of security analysis}. Region $\mathcal{R}_{12}$ follows by setting $R_{10}=0$ and replacing $Z$ with $(Z,U_1)$ and region $\mathcal{R}_{13}$ follows similarly.

\section{Examples and Applications}
In this section we present numerical examples and capacity-achieving case for our result.
 The following example shows that Regions $\mathcal{R}_1$ and $\mathcal{R}_2$ do not include each other in general. In some cases achievable points in region $\mathcal{R}_1$ are not included in region $\mathcal{R}_2$ and vice versa.
\begin{example}\label{example: R1 > R_2}
  Consider a multiple access wiretap channel where $\mathcal{X}_1=\mathcal{X}_2=\{0,1\},\mathcal{Y}=\mathcal{S}=\{0,1\}$, the channel model is described by
  \begin{equation*}
    \begin{split}
      &Y=\left\{
        \begin{aligned}
        X_1, \quad\quad\quad\quad \text{if}\;S=0,\\
        X_2, \quad\quad\quad\quad \text{if}\;S=1,
      \end{aligned}
        \right.
    \end{split}
    \;\;\;\;\;\;\;\; Z=X_2,
  \end{equation*}
  where $Pr\{S=1\}=p$. In this channel model, rate pair $(\min\{1-p,1-h(p)\},0)$ is achieved by Coding scheme 1 but cannot be achieved by Coding scheme 2 for sufficiently large $p$. If the main channel is described by 
  \begin{align*}
    Y=X_1\oplus X_2 \oplus S,
  \end{align*}
  then rate pair $(0,h(p))$ is achieved by Coding scheme 2 but cannot be achieved by Coding scheme 1 for some $p$ such that $1-h(p)\geq h(p)$.
\end{example}
The proof is given in Appendix \ref{app: proof of example 1}.
\subsection*{SD-MACs with Degraded Message Sets}
Here we consider the SD-MACs with degraded message sets, where Sender 2 sends only the common messages and Sender 1 sends both common and private messages. The private message is required to keep secret from the eavesdropper.
\begin{align*}
  &\mathcal{R}^{CSI-E}_{D,11}= \\
  &\bigcup_{P_{VUU_1U_2X_1X_2|S}}\left\{
    \begin{aligned}
      &R_1 \leq \min\{I(U_1;Y|U,U_2,V) - I(U_1;Z|U_2,U,S)\\
      &\quad\quad\quad\quad\quad\quad +R_{SK},I(U_1;Y|U,U_2,V)\};\\
      &R_0 + R_1 \leq \min\{I(V,U,U_1,U_2;Y)-I(V;S)\\
      &\quad\quad\quad\quad\quad - I(U_1;Z|U_2,U,S)+R_{SK}, \\
      &\quad\quad\quad\quad\quad I(V,U,U_1,U_2;Y)-I(V;S)\} ,
    \end{aligned}
  \right.
\end{align*}
and joint distribution such that $P_SP_{VUU_1U_2X_1X_2|S}=P_{VS}P_{UU_1U_2}P_{X_1|UU_1S}P_{X_2|UU_2S}$, where $R_{SK}= I(V;Y)-I(V;Z,U,U_2)$, and
\begin{equation*}
  \mathcal{R}^{CSI-E}_{D,12} = \bigcup_{P_{VUU_1U_2X_1X_2|S}}\left\{
    \begin{aligned}
      &R_1 \leq \min\{R_{SK},I(U_1;Y|U,U_2,V)\};\\
      &R_0 + R_1 \leq I(V,U,U_1,U_2;Y)-I(V;S).
    \end{aligned}
  \right.
\end{equation*}
Region of Coding scheme 2 in this case reduces to
\begin{align*}
  &\mathcal{R}^{CSI-E}_{D,21}= \\
  & \bigcup_{P_{UU_1U_2X_1X_2|S}}\left\{
    \begin{aligned}
      &R_1 \leq \min\left\{I(U_1;Y|U,U_2) - I(U_1;Z|U_2,U,S)\right.\\
      &\quad\quad\quad -H(S|U,U_1,U_2,Y)+H(S|Z,U,U_2),\\
   &\quad\quad\quad \left. I(U_1;Y|U,U_2)-H(S|U,U_1,U_2,Y)\right\};\\
      &R_0 + R_1 \leq \min\left\{I(U,U_1,U_2;Y) \right.\\
      &\quad\quad\quad\quad\quad\quad - I(U_1;Z|U_2,U,S)\\
      &\quad\quad\quad -H(S|U,U_1,U_2,Y)+H(S|Z,U,U_2),\\
      &\quad\quad\quad \left. I(U,U_1,U_2;Y)-H(S|U,U_1,U_2,Y)\right\}
    \end{aligned}
  \right.
\end{align*}
and joint distribution such that $P_SP_{UU_1U_2X_1X_2|S}=P_SP_{UU_1U_2}P_{X_1|UU_1S}P_{X_2|UU_2S}$,
and
\begin{equation*}
  \mathcal{R}^{CSI-E}_{D,22} = \bigcup_{P_{UU_1U_2X_1X_2|S}}\left\{
    \begin{aligned}
    &R_1 \leq \min\{H(S|Z,U,U_2)\\
    &\quad\quad\quad\quad\quad\quad -H(S|U,U_1,U_2,Y),\\
    &\quad\quad\quad\quad\quad I(U_1;Y|U,U_2)\\
    &\quad\quad\quad\quad\quad\quad -H(S|U,U_1,U_2,Y)\};\\
      &R_0 + R_1 \leq I(U,U_1,U_2;Y)\\
      &\quad\quad\quad\quad\quad\quad -H(S|U,U_1,U_2,Y).
    \end{aligned}
  \right.
\end{equation*}
The wiretap code region $\mathcal{R}^{CSI-E}_{D,3}$ is obtained by setting $R_{SK}=0$ and replacing $I(U_1;Z|U_2,U,S)$ by $I(U_1;Z|U_2,U)$ in $\mathcal{R}_1$. Let $\mathcal{R}^{CSI-E}_D$ be the convex closure of $\mathcal{R}^{CSI-E}_{D,11} \cup \mathcal{R}^{CSI-E}_{D,12}\cup\mathcal{R}^{CSI-E}_{D,21}\cup\mathcal{R}^{CSI-E}_{D,22}\cup\mathcal{R}^{CSI-E}_{D,3}$.
\begin{theorem}\label{the: inner bound of degraded message sets}
  For a multiple access wiretap channels with degraded message sets and causal state information at both encoders, it holds that $\mathcal{R}^{CSI-E}_{D} \subseteq C^{CSI-E}_{D}.$
\end{theorem}
Regions $\mathcal{R}^{CSI-E}_{D,11}$ and $\mathcal{R}^{CSI-E}_{D,12}$ follow by setting $R_{20}=R_{21}=0$ and letting $\boldsymbol{U}$ be determined by $k_{0,b}$ and common message $m_{0,b}$ in each block in Section \ref{sec: coding scheme for region r1}. Regions $\mathcal{R}^{CSI-E}_{D,21}$ and $\mathcal{R}^{CSI-E}_{D,22}$ follow by setting the private message rate of Sender 2 to 0. The details are omitted due to space limitation.
\subsubsection*{Causal CSI at One Encoder and Strictly Causal CSI at the Other}\label{sec: causal and strictly causal CSI case}
Now consider the case that causal CSI is only available at the sender who sends both messages, and strictly CSI at the other. Define region $\mathcal{R}^{CSI-SCSI}$ by setting $U_2=X_2$ in $\mathcal{R}^{CSI-E}_{D}$.
\begin{proposition}
  For SD-MAWCs with degraded messages with causal CSI at the sender that sends both messages and strictly causal CSI at the other, it holds that $\mathcal{R}^{CSI-SCSI}_D \subseteq \mathcal{C}^{CSI-SCSI}_D$.
\end{proposition}
The proof is the same as that for Theorem \ref{the: inner bound of degraded message sets} except that the input of Sender 2 is independent to the channel states. If we further assume that the wiretap channel is a degraded version of the main channel and the legitimate receiver has the knowledge of the CSI, we have the following:
\begin{theorem}\label{the: degraded mac with one-side causal CSI}
  The capacity of degraded multiple access wiretap channels with degraded message sets and causal CSI at one sender and strictly causal CSI at the other and an informed decoder is 
  \begin{align*}
    &C^{CSI-SCSI}_D = \bigcup_{P_UP_{X_1|US}P_{X_2|U}} \\
    &\left\{
      \begin{aligned}
        &R_1 \leq \min\{I(X_1;Y|U,X_2,S) - I(X_1;Z|U,X_2,S) \\
        &\quad\quad\quad\quad\quad\quad +H(S|Z,U,X_2), I(X_1;Y|U,X_2,S)\};\\
        &R_0 + R_1 \leq \min\{I(X_1,X_2;Y|S) - I(X_1;Z|U,X_2,S) \\
        &\quad\quad\quad\quad\quad\quad +H(S|Z,U,X_2), I(X_1,X_2;Y|S)\}.
      \end{aligned}
    \right.
  \end{align*}
\end{theorem}
For this model, both Coding schemes 1 and 2 are optimal. The proof is given in Appendix \ref{app: proof of theorem capacity degraded message sets}.

The capacity region in Theorem \ref{the: degraded mac with one-side causal CSI} is achieved when CSI is revealed to the decoder, and then both coding schemes are optimal. Intuitively, this can be seen as follows. By setting $V=S$, the Wyner-Ziv coding reduces to Slepian-Wolf coding. Hence, the two coding schemes mainly differ from the decoding procedure. However, by revealing CSI to the decoder, this difference is eliminated, and it is possible for both coding schemes to be optimal. The following numerical example is a `state-reproducing' channel such that the capacity is achieved.
\begin{example}\label{example: state reproduced}
  Consider a state-dependent MAC with alphabets $\mathcal{X}_1=\mathcal{X}_2=\mathcal{S}=\mathcal{Z}=\mathcal{N}=\{0,1\}$ and $\mathcal{Y}=\{0,1\}\times\{0,1\}$. The main channel output contains two components, i.e., $Y=(Y_1,Y_2)$. The channel model is described by
  \begin{align*}
   (Y_1,Y_2) = ((X_1 \odot X_2) \oplus S,X_1\odot X_2),\;\;Z = Y_2 \oplus N,
  \end{align*}
  where $S\backsim Bernoulli(q),N\backsim Bernoulli(p), q,p\in [0,\frac{1}{2}]$, and the causal CSI is available at the sender who sends both messages. The capacity of this channel is
  \begin{equation*}
    \bigcup_{\alpha\in[\frac{1}{2},1]}\left\{ 
      \begin{aligned}
        &R_1 \leq \min\{h(\alpha) + h(q) - h(p*\alpha) + h(p), h(\alpha) \},\\
        &R_1 + R_2 \leq \min\{1 + h(q) - h(\alpha*p) + h(p), 1\},
      \end{aligned}
    \right.
  \end{equation*}
  where $h(\cdot)$ is the binary entropy function with $h(a)=-a\log(a)-(1-a)\log(1-a)$.
\end{example}
The proof is given in Appendix \ref{app: proof of example 2}.



\bibliographystyle{ieeetr} 
\bibliography{ref}

\clearpage
\appendices

\section{security analysis of section \ref{sec: coding scheme for region r1}}\label{app: proof of security analysis}
The proof of the security relies on the following two lemmas.
\begin{lemma}\label{lem: wiretap codes}
  For codebooks in each block satisfying $\widetilde{R}_1-R_{10}>I(U_1;Z|U,S),\widetilde{R}_2 - R_{20} >I(U_2;Z|U,S), \widetilde{R}_1 + \widetilde{R}_2 - R_{10} - R_{20}>I(U_1,U_2;Z|S,U)$, it follows that
  \begin{align*}
    I(M_{10},M_{20};\boldsymbol{Z},\boldsymbol{S}|\textbf{C}) \leq \epsilon.
  \end{align*}
\end{lemma}
This is proved using channel resolvability \cite{helal2020cooperative} and is omitted here. The second lemma is used for secret key construction.
\begin{lemma}\label{lem: secret key}
  Suppose $V-S-(YZ)$ forms a Markov chain.
  Let $\mathcal{C}_{\mathcal{V}}=\{\boldsymbol{v}(i,j)\}_{i\in[1:2^{nR_1}],j\in[1:2^{nR_2}]}$ be a codebook containing $2^{n\widetilde{R}}$ codewords, where $\widetilde{R}=I(V;S)+\tau, R_1=I(V;S)-I(V;Y)+2\tau,R_2=I(V;Y)-\tau$, $\tau>0$. There exists a mapping $\kappa:[1:2^{nR_2}]\to\{1,...,k\}$, where $k=I(V;Y)-I(V;Z)$ such that
  \begin{align*}
    \mathrm{S}(\kappa(J)|\boldsymbol{Z},I) \leq \epsilon
  \end{align*}
  for some $\epsilon>0$ that can be arbitrarily small, where $\mathrm{S}(K|Z)=\log|\mathcal{K}| - H(K|Z)$, $\mathcal{K}$ is the range of random variable $K$. In fact, by setting the size of codebook as above, we can further construct a partition on the codebook such that
  \begin{align*}
    \mathrm{S}(I|\boldsymbol{Z}) \leq \epsilon.
  \end{align*}
\end{lemma}
When applying Lemma \ref{lem: secret key} to our model, $\boldsymbol{Z}$ is the wiretap channel output, $I$ and $J$ index the lossy description $\boldsymbol{V}(I,J)$ of state sequence $\boldsymbol{S}_{b-1}$ from the last block, $\kappa$ is the secret key mapping constructed in \emph{Key Message Codebook Generation}.
The proof is given in Appendix \ref{app: proof of lemma secret key}.
To bound the information leakage, let $\bar{\textbf{C}}^{B+1}$ be all the codebooks used in $B+1$ blocks. It follows that
\begin{align*}
  &I(M_1^{B+1},M_2^{B+1};\boldsymbol{Z}^{B+1}|\bar{\textbf{C}}^{B+1})\\
  &\leq I(M_1^{B+1},M_2^{B+1};\boldsymbol{Z}^{B+1},\boldsymbol{U}^{B+1}|\bar{\textbf{C}}^{B+1})\\
  &=\sum_{b} I(M_{1,b},M_{2,b};\boldsymbol{Z}^{B+1},\boldsymbol{U}^{B+1}|M_1^{[b+1]},M_2^{[b+1]},\bar{\textbf{C}}^{B+1})\\
  &\overset{(a)}{\leq} \sum_{b} I(M_{1,b},M_{2,b};\boldsymbol{Z}^{B+1},\boldsymbol{U}^{B+1}|\boldsymbol{S}_b,M_1^{[b+1]},M_2^{[b+1]},\bar{\textbf{C}}^{B+1})\\
  &\overset{(b)}{=}\sum_{b} I(M_{1,b},M_{2,b};\boldsymbol{Z}^{b},\boldsymbol{U}^{b}|\boldsymbol{S}_b,\bar{\textbf{C}}^{B+1}),
\end{align*}
where $(a)$ follows by the independence between $(M_{1,b},M_{2,b})$ and $\boldsymbol{S}_b$, $(b)$ follows by the independence between $(M_{1,b},M_{2,b})$ and $(M_1^{[b+1]},M_2^{[b+1]},\boldsymbol{Z}^{[b+1]},\boldsymbol{U}^{[b+1]})$ given $(\boldsymbol{Z}^b,\boldsymbol{U}^b,\boldsymbol{S}_b)$.
To bound $I(M_{1,b},M_{2,b};\boldsymbol{Z}^{b},\boldsymbol{U}^{b}|\boldsymbol{S}_{b},\bar{\textbf{C}}^{B+1})$, it follows that
\begin{align*}
  &I(M_{1,b},M_{2,b};\boldsymbol{Z}^{b},\boldsymbol{U}^{b}|\boldsymbol{S}_{b},\bar{\textbf{C}}^{B+1})\\
  &=I(M_{1,b},M_{2,b};\boldsymbol{Z}^{b-1},\boldsymbol{U}^{b-1}|\boldsymbol{S}_{b},\bar{\textbf{C}}^{B+1}) \\
  &\quad\quad\quad\quad + I(M_{1,b},M_{2,b};\boldsymbol{Z}_{b},\boldsymbol{U}_{b}|\boldsymbol{Z}^{b-1},\boldsymbol{U}^{b-1},\boldsymbol{S}_{b},\bar{\textbf{C}}^{B+1})\\
  &=I(M_{1,b},M_{2,b};\boldsymbol{Z}_{b},\boldsymbol{U}_{b}|\boldsymbol{Z}^{b-1},\boldsymbol{U}^{b-1},\boldsymbol{S}_{b},\bar{\textbf{C}}^{B+1})\\
  &=\underbrace{I(M_{10,b},M_{20,b};\boldsymbol{Z}_{b},\boldsymbol{U}_{b}|\boldsymbol{Z}^{b-1},\boldsymbol{U}^{b-1},\boldsymbol{S}_{b},\bar{\textbf{C}}^{B+1})}_{=:I_1}\\
  &+\underbrace{I(M_{11,b},M_{21,b};\boldsymbol{Z}_{b},\boldsymbol{U}_{b}|\boldsymbol{Z}^{b-1},\boldsymbol{U}^{b-1},\boldsymbol{S}_{b},M_{10,b},M_{20,b},\bar{\textbf{C}}^{B+1})}_{=:I_2},
\end{align*}
where $I_1$ satisfies
\begin{align*}
  I_1 &\leq I(M_{10,b},M_{20,b},\boldsymbol{Z}^{b-1},\boldsymbol{U}^{b-1};\boldsymbol{Z}_{b},\boldsymbol{U}_{b}|\boldsymbol{S}_{b},\bar{\textbf{C}}^{B+1})\\
  &= I(M_{10,b},M_{20,b};\boldsymbol{Z}_{b},\boldsymbol{U}_{b}|\boldsymbol{S}_{b},\bar{\textbf{C}}^{B+1}) \\
  &\quad\quad\quad\quad + I(\boldsymbol{Z}^{b-1},\boldsymbol{U}^{b-1};\boldsymbol{Z}_{b},\boldsymbol{U}_{b}|\boldsymbol{S}_{b},M_{10,b},M_{20,b},\bar{\textbf{C}}^{B+1})\\
  &\leq I(M_{10,b},M_{20,b};\boldsymbol{Z}_{b},\boldsymbol{U}_{b}|\boldsymbol{S}_{b},\bar{\textbf{C}}^{B+1}) \\
  &\quad\quad\quad\quad + I(\boldsymbol{Z}^{b-1},\boldsymbol{U}^{b-1};\boldsymbol{Z}_{b},\boldsymbol{U}_{b},\boldsymbol{S}_{b},M_{10,b},M_{20,b},K_{1,b}|\bar{\textbf{C}}^{B+1})\\
  &\overset{(a)}{=}\underbrace{I(M_{10,b},M_{20,b};\boldsymbol{Z}_{b},\boldsymbol{U}_{b}|\boldsymbol{S}_{b},\bar{\textbf{C}}^{B+1})}_{=:I_{11}} \\
  &\quad\quad\quad\quad + \underbrace{I(\boldsymbol{Z}^{b-1},\boldsymbol{U}^{b-1};K_{1,b},K_{0,b}|\bar{\textbf{C}}^{B+1})}_{=:I_{12}}.
\end{align*}
where $(a)$ follows by the Markov chain $(\boldsymbol{Z}^{b-1},\boldsymbol{U}^{b-1}) - (K_{0,b},K_{1,b},\bar{\textbf{C}}^{B+1}) - (\boldsymbol{Z}_{b},\boldsymbol{S}_{b},M_{10,b},M_{20,b})$.
By Lemma \ref{lem: wiretap codes}, we have $I_{11}\leq \epsilon$. By Lemma \ref{lem: secret key} we have
\begin{align*} 
  &I(K_{1,b};K_{0,b}|\bar{\textbf{C}}^{B+1})\leq I(K_{1,b};\boldsymbol{Z}_{b-1},\boldsymbol{U}_{b-1},K_{0,b}|\bar{\textbf{C}}^{B+1})\\
  &\quad\quad\quad\quad\quad\quad\quad \leq\mathrm{S}(K_{1,b}|\boldsymbol{Z}_{b-1},\boldsymbol{U}_{b-1},K_{0,b})\leq \epsilon,\\
  &I(K_{0,b};\boldsymbol{Z}_{b-1},\boldsymbol{U}_{b-1}|\bar{\textbf{C}^{B+1}})\leq \mathrm{S}(K_{0,b}|\boldsymbol{Z}_{b-1},\boldsymbol{U}_{b-1})\leq \epsilon.
\end{align*}
Following the same recurrence argument as in \cite{sasaki2019wiretap} yields
\begin{align*}
  I(K_{0,b},K_{1,b};\boldsymbol{Z}^{b-1},\boldsymbol{U}^{b-1}|\bar{\textbf{C}}^{B+1}) \leq 2(B+1)\epsilon
\end{align*}
and $I_1 \leq (2B+3)\epsilon$. To bound $I_{2}$, it follows that
\begin{align*}
  I_{2}&=I(M_{11,b},M_{21,b};\boldsymbol{Z}_{b},\boldsymbol{U}_b|\boldsymbol{Z}^{b-1},\boldsymbol{U}^{b-1},\boldsymbol{S}_{b},M_{10,b},M_{20,b},\bar{\textbf{C}}^{B+1})\\
  &\leq I(M_{11,b},M_{21,b},\boldsymbol{Z}^{b-1},\boldsymbol{U}^{b-1};\boldsymbol{Z}_{b},\boldsymbol{U}_b|\boldsymbol{S}_{b},M_{10,b},M_{20,b},\bar{\textbf{C}}^{B+1})\\
  &=I(M_{11,b},M_{21,b};\boldsymbol{Z}_{b},\boldsymbol{U}_b|\boldsymbol{S}_{b},M_{10,b},M_{20,b},\bar{\textbf{C}}^{B+1}) \\
  &\quad + I(\boldsymbol{Z}^{b-1},\boldsymbol{U}^{b-1};\boldsymbol{Z}_{b},\boldsymbol{U}_b|\boldsymbol{S}_{b},M_{10,b},M_{20,b},\bar{\textbf{C}}^{B+1},M_{11,b},M_{21,b})\\
  &\leq I(M_{11,b},M_{21,b};\boldsymbol{Z}_{b},\boldsymbol{U}_b,C_{11,b},C_{21,b}|\boldsymbol{S}_{b},M_{10,b},M_{20,b},\bar{\textbf{C}}^{B+1}) \\
  &\quad + I(\boldsymbol{Z}^{b-1},\boldsymbol{U}^{b-1};\boldsymbol{Z}_{b},\boldsymbol{U}_b|\boldsymbol{S}_{b},M_{10,b},M_{20,b},\bar{\textbf{C}}^{B+1},M_{11,b},M_{21,b})\\
  &=\underbrace{I(M_{11,b},M_{21,b};C_{11,b},C_{21,b}|\boldsymbol{S}_{b},M_{10,b},M_{20,b},\bar{\textbf{C}}^{B+1})}_{=:I_{21}} \\
  &\quad + \underbrace{I(M_{11,b},M_{21,b};\boldsymbol{Z}_{b},\boldsymbol{U}_b|\boldsymbol{S}_{b},M_{10,b},M_{20,b},\bar{\textbf{C}}^{B+1},C_{11,b},C_{21,b})}_{=:I_{22}}\\
  &\quad +\underbrace{I(\boldsymbol{Z}^{b-1},\boldsymbol{U}^{b-1};\boldsymbol{Z}_{b},\boldsymbol{U}_b|\boldsymbol{S}_{b},M_{10,b},M_{20,b},\bar{\textbf{C}}^{B+1},M_{11,b},M_{21,b})}_{=:I_{23}}.
\end{align*}
Then, it follows that
\begin{align*}
  &I_{21}\\
  &= I(M_{11,b},M_{21,b};C_{11,b},C_{21,b}|\boldsymbol{S}_{b},M_{10,b},M_{20,b},\bar{\textbf{C}}^{B+1})\\
  &=I(M_{11,b},M_{21,b};C_{11,b},C_{21,b}|\bar{\textbf{C}}^{B+1})\\
  &= H(C_{11,b},C_{21,b}|\bar{\textbf{C}}^{B+1}) - H(C_{11,b},C_{21,b}|\bar{\textbf{C}}^{B+1},M_{11,b},M_{21,b})\\
  &\overset{(a)}{\leq} H(C_{11,b},C_{21,b}) - H(K_{11,b},K_{21,b}|\bar{\textbf{C}}^{B+1})\\
  &\overset{(b)}{\leq} H(C_{11,b},C_{21,b}) - H(K_{11,b},K_{21,b}|K_{0,b})\\
  &\overset{(c)}{=}D(P_{K_1|K_0} || P_C) \leq \mathrm{S}(K_{1,b}|\boldsymbol{Z},K_{0,b})\leq \epsilon,
\end{align*}
where $(a)$ follows from the fact that $C_{i1,b}=M_{i1,b}\oplus K_{i1,b}$,and $M_{i1,b}$ is independent of $K_{i1,b},i=1,2$, $(b)$ follows by the Markov chain $K_{1,b}-K_{0,b}-\bar{\textbf{C}}^{B+1}$ and $K_{1,b}=(K_{11,b},K_{21,b})$, $(c)$ follows by the fact that $(C_{11,b},C_{21,b})$ is uniformly distributed on $[1:2^{nR_{11}}]\times[1:2^{nR_{21}}]$ and $P_C$ is a uniform distribution on $[1:2^{nR_{11}}]\times[1:2^{nR_{21}}]$. Bounds of $I_{22}$ and $I_{23}$ are similar to the single-user case in \cite{sasaki2019wiretap} and is omitted here. This completes the proof.

\section{proof of lemma \ref{lem: secret key}}\label{app: proof of lemma secret key}
In this section we prove Lemma \ref{lem: secret key}. The proof is divided into two steps. In Step 1, we prove there exists an equal partition on $\mathcal{C}_\mathcal{V}=\{\mathcal{C}_\mathcal{V}(k_0)\}_{k_0=1}^k$, as required by the Wyner-Ziv Theorem, such that $I(K_0;\boldsymbol{Z})\leq\epsilon$. In Step 2, based on the partition constructed in Step 1, we further construct a secret key mapping such that the constructed secret key $K_1$ satisfies
\begin{align*}
  \mathrm{S}(K_1|\boldsymbol{Z},K_0) \leq \epsilon.
\end{align*}

\subsubsection*{Step 1}
The constructions of the partition and the secret key are based on the following extractor lemma by Csisz\'ar and K\"orner \cite{csiszar2011information}.
\begin{lemma}[Lemma 17.3 in \cite{csiszar2011information}]\label{lem:extractor}
  Let $P$ be a distribution on a finite set $\mathcal{V}$ and $\mathcal{F}$ be a subset of $\mathcal{V}$ such that $\mathcal{F}=\{v\in\mathcal{V}:  P(v)\leq 1/d\}$.
For some positive number $\epsilon$, if $P(\mathcal{F})\geq 1- \epsilon$, then there exists a randomly selected mapping $G: \mathcal{V}\to\{1,\dots,k\}$ satisfies
\begin{align}
  \label{ineq:extractor}\sum_{m=1}^{k}\left| P(G^{-1}(m))-\frac{1}{k} \right| \leq 3\epsilon
 \end{align}
with probability at least $1-2ke^{-\epsilon^2(1-\epsilon)d/2k(1+\epsilon)}$.

Moreover, if each $P$ in a family $\mathcal{P}$ satisfies the hypothesis, then the probability that formula \eqref{ineq:extractor} holds for all $P\in\mathcal{P}$ is at least $1-2k|\mathcal{P}|e^{-\epsilon^2(1-\epsilon)d/2k(1+\epsilon)}$.

Thus, the desired mapping exists if $k\log k < \frac{\epsilon^2(1-\epsilon)d\log e}{2(1+\epsilon)\log 2|\mathcal{P}|}$. This realization of $G$ is denoted by $g$.
\end{lemma}
The proof in this step uses the technique in the proof of Lemma 17.5 and Theorem 17.21 in \cite{csiszar2011information} where the key idea is to construct set $\mathcal{F}$ and distribution family $\mathcal{P}$ satisfying conditions in Lemma \ref{lem:extractor}. We set parameters as follows.
\begin{equation}
  \begin{split}
    d = I(V;S) - I(V;Y),k = I(V;S) - I(V;Y) - \tau \\
  \mathcal{P} = P_{V^n} \bigcup \{P^n_{V|\boldsymbol{z}}:\boldsymbol{z}\notin \mathcal{E} \},
  \end{split}
\end{equation}
where $\mathcal{E}$ is a subset of $\mathcal{Z}^n$ with exponentially small probability and will be defined later, and $P_{V^n}$ is a uniform distribution on the codebook $\mathcal{C}_{\mathcal{V}}$.
Let $\sigma,\delta,\zeta$ be positive real numbers such that $\zeta<\delta<\sigma$. Define set
\begin{align*}
  \mathcal{T}_1 = \{\boldsymbol{s}: T^n_{P_{SV},\delta}[\boldsymbol{s}]\neq\emptyset\}
\end{align*}
and let $f$ be a function on $\mathcal{T}_1$ such that $(f(\boldsymbol{S}),\boldsymbol{S})\in T_{P_{SV},\delta}, f(\boldsymbol{S})\in\mathcal{C}_{\mathcal{V}}$. Further extend $f$ to function on $\mathcal{S}^n$ by setting $f(\boldsymbol{s})$ to some fixed sequence in $\mathcal{V}^n$ for $\boldsymbol{s}\notin\mathcal{T}_1$. By \cite[Corollary 17.9A]{csiszar2011information}, such a function $f$ always exists by setting the size of $|\mathcal{C}_{\mathcal{V}}|=2^{n(I(V;S)+\tau)}$. Define set
\begin{align*}
  \mathcal{T}_2 = \{(\boldsymbol{s},\boldsymbol{z}):\boldsymbol{s}\in \mathcal{T}_1,(f(\boldsymbol{s}),\boldsymbol{s},\boldsymbol{z})\in T^n_{P_{VSZ},\sigma}\}
\end{align*}
and let $\chi$ be the indicator function of set $\mathcal{T}_2$. Then, the joint distribution of $(\boldsymbol{v},\boldsymbol{s},\mu)$ is given by 
\begin{align}
  &P(\boldsymbol{v},\boldsymbol{z},\mu) = Pr\{f(\boldsymbol{S})=\boldsymbol{v},\boldsymbol{Z}=\boldsymbol{z},\chi(\boldsymbol{S},\boldsymbol{Z})=\mu\}\notag\\
  \label{neq: secret key lemma proof 1}&=\sum_{\boldsymbol{s}:f(\boldsymbol{s})=\boldsymbol{v},\chi(\boldsymbol{s},\boldsymbol{z})=\mu}P^n_{SZ}(\boldsymbol{s},\boldsymbol{z})
\end{align}
Define set $\mathcal{B}=\{(\boldsymbol{v},\boldsymbol{z},1):\boldsymbol{v}\in\mathcal{C}_{\mathcal{V}},\boldsymbol{z}\in T^n_{P_Z,\zeta},T^n_{P_{VSZ},\sigma}[\boldsymbol{v},\boldsymbol{z}]\neq\emptyset\}$. It follows that $P(\mathcal{B})\geq P^n_{SZ}(\mathcal{T}_2) - P^n_{Z}((T^n_{P_Z,\zeta})^c)\geq 1 - \eta^2$ for some exponentially small number $\eta$ and 
\begin{align*}
  |\mathcal{B}| &\overset{(a)}{\leq} \sum_{\boldsymbol{z}\in T^n_{P_Z,\zeta}} \left|\{\boldsymbol{v}:\boldsymbol{v}\in T^n_{P_{VZ},\sigma|\mathcal{S}|}[\boldsymbol{z}]\}\right|\\
  &\leq 2^{n(H(Z)+\epsilon)}2^{n(I(V;S) + \tau -I(V;Z) + \epsilon)},
\end{align*}
where $(a)$ follows by \cite[Lemma 2.10]{csiszar2011information}.
For any $(\boldsymbol{v},\boldsymbol{z},1)\in\mathcal{B}$, by \eqref{neq: secret key lemma proof 1} and the definition of $\mathcal{T}_2$ we have
\begin{align*}
  P(\boldsymbol{v},\boldsymbol{z},1) &\leq \sum_{\boldsymbol{s}\in T^n_{P_{VSZ},\sigma}[\boldsymbol{v},\boldsymbol{z}]} P^n_{SZ}(\boldsymbol{s},\boldsymbol{z})\\
  &\leq 2^{n(H(S|VZ)+\epsilon)}2^{-n(H(SZ)-\epsilon)}<\frac{1}{\alpha|\mathcal{B}|},
\end{align*}
where $\alpha=2^{-n(5\epsilon + \tau)}$. Define $\mathcal{B}_{\boldsymbol{z},1}:= \{\boldsymbol{v} : (\boldsymbol{v},\boldsymbol{z},1)\in\mathcal{B}\}$. By \cite[(17.15)]{csiszar2011information}, $\boldsymbol{z}\in T^n_{P_Z,\zeta}$ implies $T^n_{P_{VZ},\delta}[\boldsymbol{z}]\neq \emptyset$, and $\boldsymbol{v}\in T^n_{P_{VZ},\delta}[\boldsymbol{z}]$ is a sufficient condition of $T^n_{P_{VSZ},\sigma}[\boldsymbol{v},\boldsymbol{z}]\neq\emptyset$. Thus, the size of $\mathcal{B}_{\boldsymbol{z},1}$ can be lower bounded by
\begin{align*}
  \left| \mathcal{B}_{\boldsymbol{z},1} \right| \geq \left| \{\boldsymbol{v}: \boldsymbol{v}\in T^n_{P_{VZ},\delta}[\boldsymbol{z}],\boldsymbol{v}\in\mathcal{C}_{\mathcal{V}}\} \right| \geq 2^{n(I(V;S)-I(V;Z)+\tau-\epsilon)}.
\end{align*}
Now define $\mathcal{D}:= \{\boldsymbol{z}: P_{Z}^n(\boldsymbol{z})\geq \frac{\alpha^2|\mathcal{B}_{\boldsymbol{z},1}|}{|\mathcal{B}|}\}$ and $\mathcal{B}':= \mathcal{B}\bigcap \{\mathcal{C}_{\mathcal{V}} \times \mathcal{D}\}$. By our assumption that $I(V;Y)-I(V;Z)>\tau_1>2\tau+16\epsilon>0$, for any $(\boldsymbol{v},\boldsymbol{z},1)\in \mathcal{B}'$ with $1$ being the value of the indicator function $\chi$,
\begin{align*}
  P_{\boldsymbol{V}|\boldsymbol{Z},1}(\boldsymbol{v}|\boldsymbol{z},1)= \frac{P(\boldsymbol{v},\boldsymbol{z},1)}{P_{\boldsymbol{Z},1}(\boldsymbol{z},1)}\leq \frac{1}{\alpha^3\min|\mathcal{B}_{\boldsymbol{z},1}|} < \frac{1}{d}
\end{align*}
and
\begin{align*}
  P_{\boldsymbol{V}\boldsymbol{Z},1}(\mathcal{B}')\geq P_{\boldsymbol{V}\boldsymbol{Z},1}(\mathcal{B})-P_{\boldsymbol{Z},1}(\mathcal{D}^c) \geq P_{\boldsymbol{V}\boldsymbol{Z},1}(\mathcal{B})-\alpha^2 \geq 1-\eta_1^2
\end{align*}
for exponentially small number $\eta_1$. Now for each $\boldsymbol{z}$, denote $\mathcal{B}_{\boldsymbol{z},1}'$ the set $\{\boldsymbol{v}:(\boldsymbol{v},\boldsymbol{z},1)\in \mathcal{B}'\}$ and $\mathcal{E}\triangleq\{(\boldsymbol{z},1): P_{\boldsymbol{V}|\boldsymbol{Z},1}(\mathcal{B}_{\boldsymbol{z},1}'|\boldsymbol{z},1)\leq 1-\eta_1\}$, we have $P_{\boldsymbol{Z},1}(\mathcal{E})<\eta_1$. For each $\boldsymbol{z}\notin\mathcal{E}$, the distribution $P_{\boldsymbol{V}|\boldsymbol{Z},1}(\cdot|\boldsymbol{z},1)$ satisfies the condition in Lemma \ref{lem:extractor} with setting $\mathcal{F}=\mathcal{B}_{\boldsymbol{z},1}'(\mathcal{F} \text{ is defined in Lemma \ref{lem:extractor}})$ and hence there exists a mapping satisfying \eqref{ineq:extractor} for all $P_{\boldsymbol{V}|\boldsymbol{Z},1}(\cdot|\boldsymbol{z},1),(\boldsymbol{z},1)\notin \mathcal{E}$.

For the uniform distribution, we define the set $\mathcal{F}=\mathcal{C}_{\mathcal{V}}$. Since $P_{\boldsymbol{V}}(\boldsymbol{v})=\frac{1}{|\mathcal{C}_{\mathcal{V}|}}<\frac{1}{d}$ for any $\boldsymbol{v}\in\mathcal{C}_{\mathcal{V}}$ and $P_{\boldsymbol{V}}(\mathcal{C}_{\mathcal{V}})=1$, the conditions in Lemma \ref{lem:extractor} are also satisfied. The constructed mapping satisfies
\begin{align}
  \label{ineq: partition 1}&\sum_{m=1}^{k}\left| P_{\boldsymbol{V}}(g^{-1}(m))-\frac{1}{k} \right| \leq 3\varepsilon,\\
  \label{ineq: partition 2}&\sum_{m=1}^{k}\left| P_{\boldsymbol{V}|\boldsymbol{Z},1}(g^{-1}(m)|\boldsymbol{z},1)-\frac{1}{k} \right| \leq 3\varepsilon \;\; \text{for $(\boldsymbol{z},1)\notin\mathcal{E}$}
\end{align}
for some exponentially small number $\varepsilon$. By \eqref{ineq: partition 2} and the definition of the secure index $\mathrm{S}$, it follows that
\begin{align}
  \label{ineq: secret result of wyner-ziv index}I(g(\boldsymbol{V});\boldsymbol{Z})\leq\mathrm{S}(g(\boldsymbol{V})|\boldsymbol{Z}) \leq \varepsilon'
\end{align}
for some exponentially small number $\varepsilon'$.
The partition on the codebook arises from the mapping $g$. A codeword $\boldsymbol{v}$ belongs to bin $k$ if $g(\boldsymbol{v})=k$. Notice here by \eqref{ineq: partition 1}, the partition is not necessarily an equi-partition. Now let $g(\boldsymbol{V})$ be a random variable on $\mathcal{C}_{\mathcal{V}}$ following a nearly uniform distribution defined by the mapping $g$.

\begin{lemma}{\cite[Lemma 4]{he2016strong}}
  For any given codebook $\mathcal{C}$, if the function $g: \mathcal{C} \to [1:k]$ satisfies (\ref{ineq: partition 1}), there exists a partition  $\{\mathcal{C}_{m}\}_{m=1}^{k}$on $\mathcal{C}$ such that
  \begin{enumerate}
    \item \(|\mathcal{C}_{m}| = \frac{|\mathcal{C}|}{k}\) for all \(m \in [1: k]\), 
    \item \(H(K_0|g(\boldsymbol{V})) < 4\sqrt{\epsilon} \log{k}\),
  \end{enumerate}
where $K_0=g^{equal}(\boldsymbol{V})$ is the index of the bin containing $\boldsymbol{V}$, and $g^{equal}$ is the new mapping inducing the equal partition.
\end{lemma}
Now according to \eqref{ineq: secret result of wyner-ziv index} and the above lemma, we have
\begin{align*}
  I(K_0;\boldsymbol{Z}) &\leq I(K_0,g(\boldsymbol{V});\boldsymbol{Z})\\
  &=I(g(\boldsymbol{V});\boldsymbol{Z}) + I(K_0;\boldsymbol{Z}|g(\boldsymbol{V}))\\
  &=I(g(\boldsymbol{V});\boldsymbol{Z}) + H(K_0|g(\boldsymbol{V}))\\
  &\leq 3\varepsilon + 4\sqrt{\varepsilon} \log{k}.
\end{align*}
Note that $\varepsilon$ is an exponentially small number, and hence, the information leakage is also exponentially small.

\subsubsection*{Step 2} The proof in Step 2 is almost the same as that in Step 1 and is in fact the direct part proof of Theorem 17.21 in \cite{csiszar2011information}. Here we replace the function $f$ in Step 1 by a pair of functions $\phi_1 \times \phi_2: \mathcal{S}\to \mathcal{K}_0 \times \mathcal{K}_1$ on $\mathcal{T}_1$ such that $(\boldsymbol{V}(\phi_1(\boldsymbol{S}),\phi_2(\boldsymbol{S})),\boldsymbol{S})\in T^n_{P_{VS},\delta}$, where $\phi_1(\boldsymbol{S})=g^{equal}(f(\boldsymbol{S}))$ and $g^{equal}$ is the final mapping we construct in Step 1. The result of $\phi_2$ is the index of $\boldsymbol{V}$ in sub-bin $\mathcal{C}_{\mathcal{V}}(\phi_1(\boldsymbol{S}))$. Rewriting the set $\mathcal{B}$ as $\mathcal{B}=\{(k_0,k_1,\boldsymbol{z},1): k_0\in\mathcal{K}_0,k_1\in\mathcal{K}_1,\boldsymbol{z}\in T^n_{P_Z,\zeta},T^n_{P_{VSZ},\sigma}[\boldsymbol{v}(k_0,k_1),\boldsymbol{z}]\neq\emptyset\}$ and set $\mathcal{B}_{\boldsymbol{z},1}$ as $\mathcal{B}_{k_0,\boldsymbol{z},1}=\{k_1: (k_0,k_1,\boldsymbol{z},1)\in\mathcal{B}\}$, the remaining proof is the same as Step 1 and can be found in \cite[Proof of Theorem 17.21]{csiszar2011information}. The finally constructed mapping $\kappa$ satisfies
\begin{align*}
  \mathrm{S}(\kappa(\phi_2(\boldsymbol{S}))|\boldsymbol{Z},\phi_1(\boldsymbol{S}))\leq\epsilon
\end{align*}
and the proof is completed.

\section{proof of example \ref{example: R1 > R_2}}\label{app: proof of example 1}
Setting $p=Pr\{S=1\}$. We first consider region $\mathcal{R}_1$. Set variables as follows,
  \begin{align*}
    V=S, U_2 = \emptyset, X_1=U_1, X_2=U
  \end{align*}
with $(U,U_1)$ and $S$ being independent and $U \backsim Bernoulli(\frac{1}{2})$ and $U_1 \backsim Bernoulli(\frac{1}{2})$. It follows that $I(U_1;Z|S,U)=I(U_1;Z|S,U,U_2)=I(U_1,U_2;Z|S,U)=0$. We further have
\begin{align*}
  &I(U_1;Y|V,U,U_2)=I(U_1;Y|S,U)= H(Y|S,U) \\
  &= Pr\{S=1\}H(Y|U,S=1) = 1-p,\\
  &I(V,U,U_1,U_2;Y)-I(V;S) = H(Y)- H(S) = 1-h(p),
\end{align*}
where $h(p)=-p\log{p}-(1-p)\log(1-p)$. Note that the mutual information term $I(V;Y)=I(S;Y)\geq 0$. Hence, a rate pair $(R_1,0)$ satisfying $R_1 = \min\{1-p,1-h(p)\}$ is achievable. Now consider region $\mathcal{R}_2$. Restricting $R_2=0$ and assigning all secret key rate to Sender 1, it follows that
\begin{align*}
  &I(U_1;Z|U_2,S)= I(U_1;X_2|U_2,S)=0, \\
  &H(S|Z)=H(S|Z,U_1)=H(S),
\end{align*}
and the maximum rate can be achieved by Sender 1 will not be greater than $I(U_1;Y|U_2)$.
Hence, it is sufficient to prove $I(U_1;Y|U_2)<\min\{1-p,1-h(p)\}$, where $I(U_1;Y|U_2)$ is in fact the constraint on $R_1$ in the achievable rate region of MAC with causal CSI using Shannon strategy. To see this, note that $I(U_1;Y|U_2)$ is a convex function of conditional distribution $P_{Y|U_1U_2}$, where $P_{Y|U_1U_2}(y|u_1,u_2)=\sum_{x_1,x_2,s}P_{Y|X_1X_2S}(y|x_1,x_2,s)P_{X_1|U_1S}(x_1|u_1,s)$ $P_{X_2|U_2S}(x_2|u_2,s)P_S(s)$ is a linear function of $P_{X_1|U_1S}$ and $P_{X_2|U_2S}$. Hence, it is also a convex function of $P_{X_1|U_1S}$ and $P_{X_2|U_2S}$, where the maximum is achieved in extreme points. Thus there is no loss of generality to replace the distribution $P_{X_1|U_1S}$ and $P_{X_2|U_2S}$ with deterministic functions $x_1(u_1,s)$ and $x_2(u_2,s)$, respectively. It is shown by Example 4 in \cite{lapidoth2012multiple} that $I(U_1;Y|U_2)<\min\{1-p,1-h(p)\}$ holds when $p$ is sufficiently large.

For the second channel model, consider the case that $R_1=0$. For Coding scheme 1, it follows that 
\begin{align*}
  I(U_2;Z|U,S) = I(U_2;X_2|U,S) \overset{(a)}{=} I(U_2;X_2|U,S,U_1) 
\end{align*}
where $(a)$ follows by the fact that $(U_2,X_2)$ is independent to $U_1$ given $U$. We further have 
\begin{align*}
  I(U_2;Y|V,U,U_1) &= H(U_2|V,U,U_1) - H(U_2|V,U,U_1,Y)\\
  &\overset{(a)}{\leq} H(U_2|S,U,U_1) - H(U_2|V,U,U_1,Y,S)\\
  &\overset{(b)}{=}H(U_2|S,U,U_1) - H(U_2|U,U_1,Y,S) \\
  &\overset{(c)}{=} I(U_2;Y|S,U,U_1)\\
  &=I(U_2;X_1+X_2|S,U,U_1),
\end{align*}
where $(a)$ follows by the fact that  $U_2$ is independent to $S$ given $U$ and conditions decrease entropy, $(b)$ follows from the fact that $V$ is independent to anything else given $S$, $(c)$ follows by substituting the channel model into the mutual information. Applying the chain rule of mutual information yields
\begin{align*}
  &I(U_2;X_1+X_2,X_2|S,U,U_1)\\
   &= I(U_2;X_2|S,U,U_1) + I(U_2;X_1+X_2|S,U,U_1,X_2)\\
  &=I(U_2;X_2|S,U,U_1) + I(U_2;X_1|S,U,U_1,X_2)\\
  &\overset{(a)}{=}I(U_2;X_2|S,U,U_1)\\
  &=I(U_2;X_1+X_2|S,U,U_1) + I(U_2;X_2|S,U,U_1,X_1+X_2)
\end{align*}
where $(a)$ holds since $U_2$ is independent to $X_1$ given $(S,U,U_1,X_2)$. By the nonnegativity of the mutual information, we have $I(U_2;Z|U,S) =I(U_2;X_2|S,U,U_1)\geq I(U_2;X_1+X_2|S,U,U_1)=I(U_2;Y|S,U,U_1)$. Hence, region $\mathcal{R}_1$ reduces to $\mathcal{R}_{13}$. Now assign all the secret key rate to Sender 2. The achievable rate of Sender 2 satisfies
\begin{align*}
  R_2 &\leq I(V;Y) - I(V;Z,U,U_2) \\
  &=I(V;Y)-I(V;X_2,U,U_2)\\
  &\overset{(a)}{\leq}I(V;Y)\overset{(b)}{\leq} I(S;Y) = H(S) - H(S|Y)\overset{(c)}{\leq} h(p),
\end{align*}
where $(a)$ follows by the nonnegativity of mutual information, $(b)$ follows by data process inequality, the equality in $(c)$ holds when $V=S$ and $S$ can be determined by $Y$, which means $X_1+X_2$ takes some fixed numbers. Now considering the case that $V=S$ and $X_1+X_2$ are fixed. Without loss generality, assume $X_1+X_2=0$. The second constraint on Sender 2 in Coding scheme 1 is 
\begin{align*}
  I(U_2;Y|S,U,U_1) = H(Y|S,U,U_1) - H(Y|S,U,U_1,U_2) \overset{(a)}{=}0,
\end{align*}
where $(a)$ holds since $Y$ is determined by $S$ when $X_1+X_2=0$ is fixed. Hence, the rate of Sender 2 is $R_2=0$ in this case. We conclude that $R_2 < h(p)$ using Coding scheme 1 for this model.

For Coding scheme 2, setting $X_1=U_1,X_2=U_2, U_1 \backsim Bernoulli(\frac{1}{2})$ and $U_2 \backsim Bernoulli(\frac{1}{2})$, the first constraint on Sender 2 is 
\begin{align*}
  I(U_2;Y|U_1)&=I(X_2;Y|X_1)=H(Y|X_1) - H(Y|X_1,X_2)\\
  &= 1 - h(p).
\end{align*}
The secret key rate of Sender 2 in this case is 
\begin{align*}
  H(S|Z,U_2) - H(S|Y,U_1,U_2) &= H(S) - H(S|Y,X_1,X_2) = h(p).
\end{align*}
When the distribution of channel states satisfies $1-h(p) \geq h(p)$, we have $(0,h(p))$ can be achieved by Coding scheme 2 but cannot be achieved by Coding scheme 1.

\section{proof of theorem \ref{the: degraded mac with one-side causal CSI}}\label{app: proof of theorem capacity degraded message sets}
To see this, we first prove the achievability of Coding scheme 1. Setting  $V=S$ and $Y=(Y,S)$ in region $\mathcal{R}^{CSI-SCSI}_D$, the constraint on $R_1$ by Coding scheme 1 is
  \begin{align*}
    R_1 &\leq I(U_1;Y,S|U,X_2,S) - I(U_1;Z|U,X_2,S) + H(S|Z,U,X_2)\\
    &=I(U_1;Y|U,X_2,S) - I(U_1;Z|U,X_2,S) + H(S|Z,U,X_2)
  \end{align*}
  and $R_1 \leq I(U_1;Y|U,X_2,S)$. Now by strong functional representation lemma \cite{li2018strong}, for any random variables $(X_1,S,U)$ with conditional distribution $P_{X_1S|U}$, one can construct a random variable $U_1$ such that $X_1$ can be specified by a deterministic function $x_1(u,u_1,s)$ and $U_1$ is independent of $S$ given $U$. Together with the fact that $S$ is independent of $U$, we have
  \begin{align*}
    I(U_1;Y|U,X_2,S) \overset{(a)}{=} I(U_1,X_1;Y|U,X_2,S)\hspace{-0.05in} \overset{(b)}{=}\hspace{-0.05in} I(X_1;Y|U,X_2,S),
  \end{align*}
where $(a)$ follows by the deterministic function $x_1(u,u_1,s)$ and $(b)$ follows by the Markov chain $(U,U_1)-(X_1,X_2,S)-(Y,Z)$. Term $I(U_1;Z|U,X_2,S)$ follows similarly.
The sum rate satisfies
\begin{align*}
  &R_1 + R_2 \leq I(S,U,U_1,X_2;Y,S)-I(S;S)-I(U_1;Z|U,X_2,S) \\
  &\quad\quad\quad\quad\quad\quad\quad\quad\quad\quad\quad\quad + H(S|Z,U,X_2)\\
  &=I(U,U_1,X_2;Y|S) -I(U_1;Z|U,X_2,S) + H(S|Z,U,X_2).
\end{align*}
Applying strong functional representation lemma again completes the proof. For Coding scheme 2, we have 
\begin{align}
  R_1 \leq &I(U_1;Y|U,X_2) - I(U_1;Z|X_2,U,S) - H(S|U,U_1,X_2,Y) \notag \\
  &\quad\quad\quad\quad\quad\quad\quad\quad\quad\quad\quad\quad + H(S|Z,U,X_2)\notag\\
  &\overset{(a)}{=}I(U_1;Y,S|U,X_2) - I(U_1;S|U,X_2,Y) \notag \\
  &\quad\quad\quad\quad\quad\quad - I(U_1;Z,S|X_2,U)\notag \\
  &\quad\quad\quad\quad\quad\quad - H(S|U,U_1,X_2,Y) + H(S|Z,U,X_2)\notag\\
  &=I(U_1;Y|U,X_2,S) - I(U_1;Z|U,X_2,S) - H(S|U,X_2,Y)\notag \\
  &\quad\quad\quad\quad\quad\quad\quad\quad\quad\quad\quad\quad + H(S|U,X_2,Z),\notag\\
  \label{eq: example 2}&=I(U_1,S;Y|U,X_2) - I(U_1,S;Z|U,X_2)
\end{align}
where $(a)$ follows by the strictly causal assumption and the fact that $S$ is independent of $(U,U_1,X_2)$. Replacing $Y$ with $(Y,S)$ and applying functional representation lemma (FRL)\cite{el2011network} gives the desired bound. The rest of the proof follows similarly and is omitted here. To prove the converse, it follows that
\begin{align}
  nR_1 &= H(M_1) \notag \\
  &\leq I(M_1;\boldsymbol{Y},\boldsymbol{S}|M_0,\boldsymbol{X}_2) - I(M_1;\boldsymbol{Z}|M_0,\boldsymbol{X}_2) + \delta \notag\\
  &\overset{(a)}{=}I(M_1;\boldsymbol{Y},\boldsymbol{S}|\boldsymbol{Z},M_0,\boldsymbol{X}_2) + \delta \notag\\
  &\leq I(M_1,\boldsymbol{X}_1,\boldsymbol{S};\boldsymbol{Y},\boldsymbol{S}|\boldsymbol{Z},M_0,\boldsymbol{X}_2) + \delta \notag\\
  &= I(\boldsymbol{X}_1,\boldsymbol{S};\boldsymbol{Y},\boldsymbol{S}|\boldsymbol{Z},M_0,\boldsymbol{X}_2)\notag \\
  &\quad\quad\quad\quad + I(M_1;\boldsymbol{Y},\boldsymbol{S}|\boldsymbol{Z},M_0,\boldsymbol{X}_2,\boldsymbol{X}_1,\boldsymbol{S}) + \delta \notag\\
  &\overset{(b)}{=} I(\boldsymbol{X}_1,\boldsymbol{S};\boldsymbol{Y},\boldsymbol{S}|\boldsymbol{Z},M_0,\boldsymbol{X}_2) + \delta \notag\\
  &=H(\boldsymbol{X}_1,\boldsymbol{S}|\boldsymbol{Z},M_0,\boldsymbol{X}_2) - H(\boldsymbol{X}_1,\boldsymbol{S}|\boldsymbol{Z},M_0,\boldsymbol{X}_2,\boldsymbol{Y},\boldsymbol{S}) + \delta \notag\\
  &\overset{(c)}{=}H(\boldsymbol{X}_1,\boldsymbol{S}|\boldsymbol{Z},M_0,\boldsymbol{X}_2) - H(\boldsymbol{X}_1,\boldsymbol{S}|M_0,\boldsymbol{X}_2,\boldsymbol{Y},\boldsymbol{S}) + \delta \notag\\
  \label{neq: R1 upper bound 1}&=I(\boldsymbol{X}_1,\boldsymbol{S};\boldsymbol{Y},\boldsymbol{S}|M_0,\boldsymbol{X}_2) - I(\boldsymbol{X}_1,\boldsymbol{S};\boldsymbol{Z}|M_0,\boldsymbol{X}_2) + \delta
\end{align}
where $(a),(c)$ follows by the degradedness of the wiretap channel, $(b)$ follows from the Markov chain $(M_1,M_0)-(\boldsymbol{X}_1,\boldsymbol{X}_2,\boldsymbol{S})-(\boldsymbol{Y},\boldsymbol{Z})$. Set $U_i=(Y^{i-1},S^{i-1},\boldsymbol{X}_2,M_0)$, it follows that $U_i$ is independent of $S_i$. Then, we have
\begin{align*}
  &I(\boldsymbol{X}_1,\boldsymbol{S};\boldsymbol{Y},\boldsymbol{S}|M_0,\boldsymbol{X}_2) - I(\boldsymbol{X}_1,\boldsymbol{S};\boldsymbol{Z}|M_0,\boldsymbol{X}_2)\\
  &= \sum_{i=1}^n I(\boldsymbol{X}_1,\boldsymbol{S};Y_i,S_i|Y^{i-1},S^{i-1},M_0,\boldsymbol{X}_2) \\
  &\quad\quad\quad\quad\quad\quad - I(\boldsymbol{X}_1,\boldsymbol{S};Z_i|Z^{i-1},M_0,\boldsymbol{X}_2)\\
  &=\sum_{i=1}^n H(Y_i,S_i|Y^{i-1},S^{i-1},M_0,\boldsymbol{X}_2) \\
  &\quad\quad\quad\quad - H(Y_i,S_i|Y^{i-1},S^{i-1},M_0,\boldsymbol{X}_2,\boldsymbol{X}_1,\boldsymbol{S})\\
  &\quad\quad\quad\quad - H(Z_i|Z^{i-1},M_0,\boldsymbol{X}_2) + H(Z_i|Z^{i-1},M_0,\boldsymbol{X}_2,\boldsymbol{X}_1,\boldsymbol{S})\\
  &\overset{(a)}{\leq}\sum_{i=1}^n H(Y_i,S_i|U_i,X_{2,i}) - H(Y_i,S_i|U_i,X_{2,i},X_{1,i},S_i)\\
  &\quad\quad\quad\quad - H(Z_i|Y^{i-1},S^{i-1},Z^{i-1},M_0,\boldsymbol{X}_2) \\
  &\quad\quad\quad\quad + H(Z_i|Y^{i-1},S^{i-1},M_0,\boldsymbol{X}_2,X_{1,i},S_{i})\\
  &\overset{(b)}{=}\sum_{i=1}^n H(Y_i,S_i|U_i,X_{2,i}) - H(Y_i,S_i|U_i,X_{2,i},X_{1,i},S_i)\\
  &\quad\quad\quad\quad - H(Z_i|Y^{i-1},S^{i-1},M_0,\boldsymbol{X}_2) \\
  &\quad\quad\quad\quad + H(Z_i|Y^{i-1},S^{i-1},M_0,\boldsymbol{X}_2,X_{1,i},S_{i})\\
  &=\sum_{i=1}^n I(X_{1,i},S_i;Y_i,S_i|U_i,X_{2,i}) - I(X_{1,i},S_i;Z_i|U_i,X_{2,i})\\
  &=n(I(X_1;Y,S|U,X_2,S) - I(X_1;Z|U,X_2,S)+H(S|Z,U,X_2)),
\end{align*}
where $(a)$ and $(b)$ follows by the Markov chain $(U_i,M_0)-(X_{1,i},X_{2,i},S_i)-(Y_i,Z_i)$ and the degradedness of the wiretap channel. The second constraint on $R_1$ follows similarly and is omitted here. For the sum rate, it follows that
\begin{align*}
  &n(R_0+R_1) \\
  &= H(M_0) + nR_1\\
  &\leq I(M_0;\boldsymbol{Y},\boldsymbol{S}) + nR_1 +\delta\\
  &\leq \sum_{i=1}^n I(M_0;Y_i,S_i|Y^{i-1},S^{i-1}) + I(X_{1,i},S_i;Y_i,S_i|U_i,X_{2,i}) \\
  &\quad\quad\quad\quad - I(X_{1,i},S_i;Z_i|U_i,X_{2,i}) + \delta \\
  &\leq \sum_{i=1}^n I(M_0,Y^{i-1},S^{i-1},\boldsymbol{X}_2;Y_i,S_i)  + I(X_{1,i},S_i;Y_i,S_i|U_i,X_{2,i}) \\
  &\quad\quad\quad\quad - I(X_{1,i},S_i;Z_i|U_i,X_{2,i}) + \delta \\
  &= \sum_{i=1}^n I(U_i,X_{1,i},X_{2,i},S_i;Y_i,S_i)- I(X_{1,i},S_i;Z_i|U_i,X_{2,i}) + \delta\\
  &=n(I(X_1,X_2;S)-I(X_1;Z|U,X_2,S)+ H(S|Z,U,X_2)+\delta).
\end{align*}
The second constraint on the sum rate follows similarly and is omitted here.
\section{Proof of example \ref{example: state reproduced}}\label{app: proof of example 2}
\emph{Converse: } By Appendix \ref{app: proof of theorem capacity degraded message sets} and replacing $(Y,S)$ with $Y$, it follows that (for simplicity, we omit $\delta$.)
\begin{align*}
  nR_1 &\leq I(\boldsymbol{X}_1,\boldsymbol{S};\boldsymbol{Y}|M_0,\boldsymbol{X}_2) - I(\boldsymbol{X}_1,\boldsymbol{S};\boldsymbol{Z}|M_0,\boldsymbol{X}_2)\\
  &=(H(\boldsymbol{Y}|M_0,\boldsymbol{X}_2) - H(\boldsymbol{Y}|M_0,\boldsymbol{X}_2,\boldsymbol{X}_1,\boldsymbol{S})) - (H(\boldsymbol{Z}|M_0,\boldsymbol{X}_2)\\
  &\quad\quad\quad\quad -H(\boldsymbol{Z}|M_0,\boldsymbol{X}_2,\boldsymbol{X}_1,\boldsymbol{S}))\\
  &=(H(\boldsymbol{Y}_2|M_0,\boldsymbol{X}_2) + H(\boldsymbol{Y}_1|M_0,\boldsymbol{X}_2,\boldsymbol{Y}_2)) \\
  &\quad\quad\quad\quad - (H(\boldsymbol{Z}|M_0,\boldsymbol{X}_2)-nh(p))\\
  &\overset{(a)}{\leq} n + nh(q) - (H(\boldsymbol{Y}_2 \oplus N|M_0,\boldsymbol{X}_2)-nh(p)),
\end{align*}
where $(a)$ follows by $H(\boldsymbol{Y}_2|M_0,\boldsymbol{X}_2)\leq \sum_{i=1}^n H(Y_i)\leq n$.
Hence, there exists some $\alpha \in [\frac{1}{2},1]$ such that
\begin{align*}
  nR_1 \leq nh(\alpha) + nh(q) - (H(\boldsymbol{Y}_2 \oplus N|M_0,\boldsymbol{X}_2)-nh(p)).
\end{align*}
To bound $H(\boldsymbol{Y}_2 \oplus N|M_0,\boldsymbol{X}_2)$, we invoke Lemmas 5 and 6 in \cite{liang2008multiple} (proposed in \cite{wyner1973theorem}).
\begin{align*}
  &H(\boldsymbol{Y}_2 \oplus N|M_0,\boldsymbol{X}_2)\\
  &= \mathbb{E}_{M_0,\boldsymbol{X}_2} \left[ H(\boldsymbol{Y}_2 \oplus N|M_0=m_0,\boldsymbol{X}_2=\boldsymbol{x}_2) \right] \\
  &\overset{(a)}{\geq} \mathbb{E}_{M_0,\boldsymbol{X}_2} \left[n h\left( p*h^{-1}\left( \frac{H(\boldsymbol{Y}_2|M_0=m_0,\boldsymbol{X}_2=\boldsymbol{x}_2)}{n} \right) \right) \right]\\
  &\overset{(b)}{\geq} nh\left( p*h^{-1}\left(  \mathbb{E}_{M_0,\boldsymbol{X}_2} \left[\frac{H(\boldsymbol{Y}_2|M_0=m_0,\boldsymbol{X}_2=\boldsymbol{x}_2)}{n} \right]\right) \right) \\
  &=nh\left( p*h^{-1}\left(  \frac{H(\boldsymbol{Y}_2|M_0,\boldsymbol{X}_2)}{n}\right) \right) \\
  &=nh\left( p*h^{-1}\left(  \frac{nh(\alpha)}{n}\right) \right) = nh(p*\alpha).
\end{align*}
where $(a)$ follows by Lemma 6 in \cite{liang2008multiple}, $(b)$ follows by Lemma 5 in \cite{liang2008multiple}. Combining previous bounds yields
\begin{align}
  \label{neq: numerical example upper bound R1 1}nR_1 \leq nh(\alpha) + nh(q) - nh(p*\alpha) + nh(p).
\end{align}
For the second constraint, by Appendix \ref{app: proof of theorem capacity degraded message sets}, we have
\begin{align}
  nR_1 &\leq I(\boldsymbol{X}_1;\boldsymbol{Y}|M_0,\boldsymbol{X}_2,\boldsymbol{S})\notag\\
  &=I(\boldsymbol{X}_1,\boldsymbol{S};\boldsymbol{Y}|M_0,\boldsymbol{X}_2) - I(\boldsymbol{S};\boldsymbol{Y}|M_0,\boldsymbol{X}_2)\notag\\
  &\overset{(a)}{=}I(\boldsymbol{X}_1,\boldsymbol{S};\boldsymbol{Y}|M_0,\boldsymbol{X}_2) - H(\boldsymbol{S})\notag\\
  \label{neq: numerical example upper bound R1 2}&=nh(\alpha) + nh(q) - nh(q)=nh(\alpha).
\end{align}
where $(a)$ follows by the fact that $\boldsymbol{S}$ is independent to $(M_0,\boldsymbol{X}_2)$ and the receiver can always recover $\boldsymbol{S}$ by $S_i = Y_{1,i} \ominus Y_{2,i}$.

For the first constraint of the sum rate, by Appendix \ref{app: proof of theorem capacity degraded message sets}, we have 
\begin{align}
  n(R_1 + R_2) &\leq I(M_0;\boldsymbol{Y}) + nR_1 \notag\\
  &\leq I(M_0,\boldsymbol{X}_2;\boldsymbol{Y}) + nR_1\notag\\
  &\leq I(M_0,\boldsymbol{X}_1,\boldsymbol{X}_2,\boldsymbol{S};\boldsymbol{Y}) - I(\boldsymbol{X}_1,\boldsymbol{S};\boldsymbol{Z}|M_0,\boldsymbol{X}_2)\notag \\
  \label{neq: numerical example upper bound R1+R2 1}&\leq n + nh(q) - nh(\alpha*p) + nh(p).
\end{align}
The second constraint follows similarly and is omitted.

\emph{Direct Part: } The achievable region is obtained by first applying FRL to \eqref{eq: example 2} directly, and other terms follows similarly by setting $U_2=X_2$ in region $\mathcal{R}^{CSI-E}_{D,21}$ and using FRL as \eqref{eq: example 2}.  Then, set $Pr\{X_2 = 1\}=1$ and define $X_1 = U \oplus S \oplus X'$, where $Pr\{U=1\}=\frac{1}{2}$, $X'$ is a random variable independent to $(U,X_2,S,N)$ with $Pr\{X'=1\}=\beta$ such that $q*\beta=\alpha$ for some $\alpha\in[\frac{1}{2},1]$. The proof is completed.

\end{document}